\documentclass[twocolumn,preprintnumbers,prl,superscriptaddress,amsmath,amssymb,amsfonts,longbibliography]{revtex4-2}

\usepackage{graphicx}  
\usepackage{color}
\usepackage{bm}
\usepackage[colorlinks,bookmarks=true,citecolor=blue,linkcolor=blue,urlcolor=blue, breaklinks=true]{hyperref}

\begin{document}

\title{Fully gapped pairing state in spin-triplet superconductor UTe$ _2$}

\author{S. Suetsugu}
\affiliation{Department of Physics, Kyoto University, Kyoto 606-8502, Japan}
\author{M. Shimomura}
\affiliation{Department of Physics, Kyoto University, Kyoto 606-8502, Japan}
\author{M. Kamimura}
\affiliation{Department of Physics, Kyoto University, Kyoto 606-8502, Japan}
\author{T. Asaba}
\affiliation{Department of Physics, Kyoto University, Kyoto 606-8502, Japan}
\author{H. Asaeda}
\affiliation{Department of Physics, Kyoto University, Kyoto 606-8502, Japan}
\author{Y. Kosuge}
\affiliation{Department of Physics, Kyoto University, Kyoto 606-8502, Japan}
\author{Y. Sekino}
\affiliation{Department of Physics, Kyoto University, Kyoto 606-8502, Japan}
\author{S. Ikemori}
\affiliation{Department of Physics, Kyoto University, Kyoto 606-8502, Japan}
\author{Y. Kasahara}
\affiliation{Department of Physics, Kyoto University, Kyoto 606-8502, Japan}
\author{Y. Kohsaka}
\affiliation{Department of Physics, Kyoto University, Kyoto 606-8502, Japan}
\author{M. Lee}
\affiliation{Department of Physics, University of Colorado Boulder, Boulder, CO 80309, USA}
\author{Y. Yanase}
\affiliation{Department of Physics, Kyoto University, Kyoto 606-8502, Japan}
\author{H. Sakai}
\affiliation{Advanced Science Research Center, Japan Atomic Energy Agency, Tokai, Ibaraki 319-1195, Japan}
\author{P. Opletal}
\affiliation{Advanced Science Research Center, Japan Atomic Energy Agency, Tokai, Ibaraki 319-1195, Japan}
\author{Y. Tokiwa}
\affiliation{Advanced Science Research Center, Japan Atomic Energy Agency, Tokai, Ibaraki 319-1195, Japan}
\author{Y. Haga}
\affiliation{Advanced Science Research Center, Japan Atomic Energy Agency, Tokai, Ibaraki 319-1195, Japan}
\author{Y. Matsuda}
\affiliation{Department of Physics, Kyoto University, Kyoto 606-8502, Japan}

\date{\today}

\begin{abstract}
Spin-triplet superconductors provide an ideal platform for realizing topological superconductivity with emergent Majorana quasiparticles.  The promising candidate is the recently discovered superconductor UTe$ _2$, but the symmetry of the superconducting order parameter remains highly controversial.  Here we determine the superconducting gap structure by the thermal conductivity of ultra-clean UTe$ _2$ single crystals. We find that the $a$ axis thermal conductivity divided by temperature $\kappa/T$ in zero-temperature limit is vanishingly small for both magnetic fields $\bm{H}||a$ and $\bm{H}||c$ axes up to $H/H_{c2}\sim 0.2$, demonstrating the absence of any types of nodes around $a$ axis contrary to the previous belief.  The present results, combined with the reduction of the NMR Knight shift in the superconducting state, indicate that the superconducting order parameter belongs to the isotropic $A_u$ representation with a fully gapped pairing state, analogous to the B phase of superfluid $ ^3$He. These findings reveal that UTe$ _2$ is likely to be a long-sought three-dimensional (3D) strong topological superconductor characterized by a 3D winding number, hosting helical Majorana surface states on any crystal plane.
\end{abstract}


\maketitle

\section{Introduction}
Spin-triplet pairing states have aroused tremendous interest because of the emergence of Majorana quasiparticles \cite{sato2017topological} and their potential application to fault-tolerant quantum information processing \cite{PhysRevLett.86.268,Kitaev2003}. The most famous example is the superfluid $ ^3$He \cite{RevModPhys.47.331}, and the quest for its superconducting analogue has been a long-standing issue in condensed matter physics. The promising candidate is the recently discovered heavy fermion superconductor UTe$ _2$ \cite{ran2019nearly}. UTe$ _2$ exhibits highly unusual superconducting properties, including extremely large upper critical fields well exceeding the Pauli limit \cite{ran2019nearly,aoki2019unconventional}, coexistence of superconductivity and ferromagnetic order in high magnetic field \cite{ran2019extreme,knafo2021comparison}, reentrant superconductivity that resembles to ferromagnetic superconductors URhGe and UCoGe \cite{ran2019extreme,knebel2019field}, and peculiar behavior of nuclear magnetic resonance (NMR) Knight shift and absence of coherence peak in the superconducting state \cite{ran2019nearly,nakamine2019superconducting,PhysRevB.103.L100503,fujibayashi2022superconducting}. All of these striking properties are indicative of the spin-triplet pairing state. Moreover, this unconventional superconducting state occurs in the paramagnetic state at ambient pressure. This is in contrast to the other ferromagnetic superconductors; at ambient pressure, UGe$_2$ shows a ferromagnetic order \cite{saxena2000superconductivity} and in URhGe and UCoGe, superconductivity coexists with ferromagnetic order \cite{aoki2001coexistence,PhysRevLett.99.067006}. Therefore, UTe$ _2$ is a long-sought material that allows us to examine superconducting properties in more detail using various probes.

Of particular interest is the superconducting gap structure of UTe$ _2$, which is of primary importance for understanding the peculiar superconducting state associated with the spin-triplet pairing state. Despite the intensive research efforts, however, the superconducting order parameter has been highly controversial and its determination is still challenging. At an early stage, a chiral superconducting state with a multicomponent order parameter has been suggested by the double peak of the specific heat and polar Kerr effect \cite{hayes2021multicomponent}, which was supported by scanning tunneling spectroscopy experiments \cite{jiao2020chiral}. However, as the sample quality improves, a single peak is observed in the specific heat \cite{cairns2020composition,PhysRevB.104.224501,rosa2022single} and does not split into two peaks even under uniaxial strain \cite{girod2022thermodynamic}. Moreover, recent Kerr effect experiments on clean crystals report no evidence of broken time-reversal symmetry \cite{ajeesh2023fate}. Considering the orthorhombic crystal structure of UTe$_2$ [Fig.\,1(a)], although the chiral state may appear at the surface, chiral superconductivity is unlikely to be realized in the bulk.

Given no solid evidence for the bulk chiral superconductivity in UTe$_2$, there are four possible gap symmetries for spin triplet superconductivity, $\{A_u, B_{1u},B_{2u}, B_{3u}\}$ \cite{aoki2022unconventional}. The $A_u$ symmetry is fully gapped [Fig.\,1(b)], corresponding to the B phase of superfluid $^3$He. On the other hand, $B_{1u}$, $B_{2u}$, and $B_{3u}$ symmetries have point nodes at isolated points on the Fermi surface along the $c$, $b$, and $a$ axes, respectively [Fig.\,1(c)]. Several measurements, such as thermal conductivity \cite{PhysRevB.100.220504}, penetration depth \cite{bae2021anomalous,ishihara2021chiral} and specific heat \cite{PhysRevResearch.2.032014}, have reported the presence of the low energy quasiparticle excitations, suggesting the presence of point nodes rather than line nodes. NMR measurements show that below the superconducting transition temperature $T_c$, the Knight shift along the $b$ and $c$ axes decreases \cite{PhysRevB.103.L100503,fujibayashi2022superconducting}. This excludes the $B_{1u}$ and $B_{2u}$ symmetries. Based on these results, it has been widely discussed that the $B_{3u}$ symmetry with point nodes along the $a$ axis accounts for the superconducting gap of UTe$_2$. The point node(s) around the $a$ axis has also been suggested by thermal conductivity \cite{PhysRevB.100.220504}.

However, it should be stressed that since the low energy quasiparticle excitations in spin-triplet superconductors are extremely sensitive to impurities \cite{mineev2022low}, the determination of the gap symmetry in the previous measurements may be hindered by impurity. Moreover, as mentioned previously, the gap structure near the surface may be distinct from the bulk. Therefore, it is premature to conclude the presence of point nodes along the $a$ axis. Determining whether the gap symmetry is $A_u$ or $B_{3u}$ is of crucial importance, because the topological properties of these two symmetries are fundamentally different \cite{PhysRevLett.123.217001}. Thus the situation calls for reexamining the gap structure using a bulk probe on crystals with an extremely low impurity concentration. Here, we determined the superconducting gap structure of ultra-clean UTe$_2$ crystals \cite{aoki2022first,PhysRevMaterials.6.073401} by thermal conductivity $\kappa$, which is a bulk and directional probe of the superconducting gap structure \cite{matsuda2006nodal}. In contrary to the previous reports, the present results demonstrate the absence of any type of nodes at and around the $a$ axis. This indicates that the superconducting order parameter of UTe$_2$ belongs to the isotropic $A_u$ representation with a fully gapped pairing state, analogous to the B phase of superfluid $ ^3$He.

\section{Methods}
High-quality single crystals of UTe$ _2$ were grown by a flux method as described in ref.\,\citenum{PhysRevMaterials.6.073401}. We used two crystals \#1 and \#2 from different batches. Sample \#1 was cleaved into two pieces. The data for sample \#1 are primarily from measurements on one piece. The other piece of sample \#1 was used for resistivity experiments to obtain the upper critical fields and magnetoresistance. The size of the former piece is 2511 $\mathrm{\mu m}$ (length) $\times$ 147.5 $\mathrm{\mu m}$ (width) $\times$ 77.5 $\mathrm{\mu m}$ (thickness). The size of the latter piece is 2230 $\mathrm{\mu m}$ (length) $\times$ 204.5 $\mathrm{\mu m}$ (width) $\times$ 75.3 $\mathrm{\mu m}$ (thickness). The size of sample \#2 is 2467 $\mathrm{\mu m}$ (length) $\times$ 178.8 $\mathrm{\mu m}$ (width) $\times$ 104 $\mathrm{\mu m}$ (thickness). The resistivity was measured by a standard four probe method by applying current along the crystal $a$ axis. Four gold wires were attached by spot welding. The specific heat was measured by a long-relaxation time method as described in Refs.\,\citenum{PhysRevB.63.094508} and \citenum{PhysRevLett.99.057001}.

Thermal conductivity $\kappa$ was measured by the standard steady-state method by applying temperature gradient $\bm{Q}$ along the crystal $a$ axis. To obtain good contacts with low contact resistance of $\sim$0.5\,$\Omega$, four gold wires were attached by spot welding. The gold wires serve heat links to a 10-$\mathrm{k\Omega}$ chip resistor as a heater, two field calibrated thermometers, and a silver plate. The silver plate was fixed with silver paste to a copper block as a heat bath.

\section{Results}
The superconducting transition temperature $T_c$ of both crystals \#1 and \#2 is 2.1\,K, which is higher than the previously reported value of typically $\sim$ 2.0\,K \cite{rosa2022single}. Figure\,1(d) displays the temperature ($T$) dependence of the resistivity $\rho(T)$ along the $a$ axis for sample \#2. The normal state resistivity is well fitted by $\rho(T)=\rho_0 + AT^2$ [inset of Fig.\,1(d)]. The quadratic temperature dependence of $\rho$, a characteristic of Fermi liquid, indicates the importance of the electron-electron correlation. The residual resistivity ratio ($\mathrm{RRR}\equiv\rho(300\ \mathrm{K})/\rho_0$) of 260 for sample \#1 and 630 for sample \#2 is nearly 10 and 30 times larger than that reported in crystal with $\mathrm{RRR}=22$\cite{PhysRevB.100.220504}, respectively. Hereafter, the samples with RRR = 22 \cite{PhysRevB.100.220504}, 260 (\#1), and 630 (\#2) will be referred to as samples in moderately clean, very-clean, and ultra-clean regions, respectively. It should be noted that clear quantum oscillations are reported in the sample with RRR = 220 \cite{aoki2022first}. These confirm the high quality of the present crystals. By comparing these crystals with different RRRs, we can obtain pivotal information about the quasiparticle excitations that are intimately related to the superconducting gap function.

The upper critical fields determined by the resistive transition of sample \#1 for ${\bm H}||a$ and ${\bm H}||c$, $H_{c2}^a$ and $H_{c2}^c$, respectively, are displayed in Fig.\,1(e). At zero temperature, $H_{c2}^a(0)$ and $H_{c2}^c(0)$ are approximately 12\,T and 17\,T, respectively. The upper critical fields of sample \#2 is close to those of \#1. As shown in the inset, $H_{c2}^a$ is very close to $H_{c2}^c$ in the vicinity of $T_c$. Since the initial slope at $T_c$ is proportional to the orbital limiting field, $-(dH_{c2}^{a,c}/dT)|_{T_c}\propto 1/\xi_b\xi_{c,a}$, where $\xi_a,\xi_b$, and $\xi_c$ are the coherence length along the $a$, $b$ and $c$ axes, respectively, the present results indicate $\xi_a\approx \xi_c$. This indicates that the average Fermi velocity along the $a$ axis is very close to that along the $c$ axis.

The heat capacity $C$ of sample \#1 exhibits a very sharp transition at $T_c$ with no peak splitting (Fig.\,S1 in Supplemental Material \cite{supp}). For comparison, we plot the reported data for a very-clean crystal with $T_c = 2.1$\,K \cite{aoki2022first}, whose RRR = 220 is close to sample \#1, and for a crystal with $T_c = 1.7$\,K \cite{rosa2022single}. The temperature dependence of $C/T$ for sample \#1 is very close to the data of the very clean crystal with a similar RRR that shows a negligibly small residual $C/T$ in the zero temperature limit. These results further support the high quality of the present crystals. The electronic heat capacity coefficient in the normal state $\gamma$ is 120\,mJ/K$^2$mol. The Kadowaki-Woods ratio, $A/\gamma^2\approx 2\times10^{-5}$ $\mathrm{\mu\Omega}$cm(molK/mJ)$ ^2$, is close to that of typical correlated systems \cite{kadowaki1986universal}. 

Figure\,2 shows the $T$-dependence of $\kappa / T$ with applied thermal current $\bm{Q}||a$ for the very-clean (\#1) and ultra-clean (\#2) crystals, along with $\kappa/T$ for the moderately-clean crystal \cite{PhysRevB.100.220504}. For sample \#1, $\kappa/T$ in the normal state in zero field above $T_c$ is close to the electronic contribution estimated from the Wiedemann-Franz law, $L_0 / \rho(H=0)$, (blue dashed line), where $L_0= 2.44\times10^{-8}$\,W$\mathrm{\Omega}$K$^{-2}$ is the Lorenz number. Here $\rho(H=0)$ is extrapolated below $T_c$ by $\rho(T)=\rho_0 + AT^2$. The normal state value of $\kappa/T$, $\kappa_\mathrm{N}/T$, at lower temperatures is measured above $H_{c2}^a$ (taken at $\mu_0 H =$ 12\,T), which is very close to $L_0 / \rho(H=0)$ below 0.5\,K. These results demonstrate that the thermal conductivity in the normal state is dominated by the electronic contribution in the very- and ultra-clean crystals. In contrast, $L_0 / \rho(H=0)$ for the moderately clean crystal is much smaller than $\kappa/T$ above $T_c = 1.7$\,K, indicating a dominant phonon contribution. In zero field, $\kappa/T$ exhibits a distinct kink upon entering the superconducting state, increases steeply and reaches a maximum at $\sim$1.5\,K and $\sim$1.2\,K for samples \#1 and \#2, respectively. As observed in several strongly correlated systems \cite{PhysRevLett.69.1431,PhysRevLett.87.057002,PhysRevLett.99.116402}, the enhancement of $\kappa/T$ below $T_c$ is attributed to the rapid enhancement of the quasiparticle mean free path, which is caused by the suppression of the electron-electron inelastic scattering rate due to the superconducting gap formation. The enhancement of $\kappa/T$ of sample \#2 is more significant than sample \#1 because the mean free path is larger in samples with larger RRR. We note that the enhancement of $\kappa/T$ in the moderately clean crystal is much smaller than in these crystals. Moreover, $\kappa_\mathrm{0N}/T \equiv \kappa_\mathrm{N}/T (T \rightarrow 0)$ of sample \#1 determined by the data taken at 12\,T is 1.9 W/K$ ^\mathrm{2}$m, one order of magnitude larger than 0.1 W/K$ ^\mathrm{2}$m reported for the moderately clean crystal \cite{PhysRevB.100.220504}. The difference of $\kappa_\mathrm{0N}/T$ also reflects the significantly enhanced mean free path in the very-clean and ultra-clean crystals.

Figure\,3(a) shows the $T$-dependence of $\kappa/T$ at low temperatures for samples \#1 and \#2. The thermal conductivity in the superconducting state has quasiparticle and phonon contributions, $\kappa=\kappa_{qp}+\kappa_{ph}$. In the boundary-limited scattering regime at low temperatures, $\kappa_{ph}/T$ is proportional to $\ell_{ph}T^2$, where $\ell_{ph}$ is the phonon mean free path. While $\ell_{ph}$ is $T$-independent for diffuse scattering limit, resulting in $\kappa_{ph}/T\propto T^2$, it is $T^{-1}$-dependent for specular reflection, resulting in $\kappa_{ph}/T\propto T$. In real systems, $\kappa_{ph}/T \propto T^{p}$ with $1\leq p \leq 2$. Since $\kappa_{ph}/T$ becomes zero at $T=0$, $\kappa_0/T\equiv \kappa/T (T\rightarrow 0)$ contains only the quasiparticle contribution. It is apparent the linear extrapolation of $\kappa/T$ to $T=0$ yields a negative intercept. We find that $\kappa/T$ for both samples are best fitted by $\kappa/T\propto T^{\alpha}$ with $\alpha$=1.48 below 0.3\,K [inset of Fig.\,3(a)]. These results indicate the absence of residual thermal conductivity $\kappa_0/T \approx 0$. For unconventional superconductors with line nodes in the energy gap, a finite residual term is expected due to the existence of a residual normal fluid \cite{PhysRevLett.79.483}. Therefore, the present results provide evidence for the absence of line nodes. This is further supported by the estimation of the residual term in a line-nodal superconductor. For line nodes, the universal expression of $\kappa_0 / T$ for unitary scattering is given by $\frac{L_0}{\mu_0\lambda^2} \frac{\hbar}{\pi\Delta_0}$, where $\mu_0$ is the permeability of vacuum and $\lambda$ is the penetration depth. Using $2\Delta_0 = 3.5k_BT_c$ and $\lambda \approx 1000$\,nm \cite{PhysRevB.100.220504,bae2021anomalous}, we obtain $\kappa_0/T\approx 0.054\,$W/K$ ^2$m (gray dashed line). Clearly, $\kappa/T$ at the lowest temperature is less than the calculated value for the line nodes. 

We next examine the presence/absence of point nodes. It is well known that the heat transport at low magnetic fields in nodal superconductors is fundamentally different from that in full-gap superconductors. In magnetic fields, the energy of quasiparticles with momentum $\hbar\bm{k}$ in a circulating supercurrent flow $\bm{v}_s\,$ shifts as $E_{\bm{k}} \rightarrow E_{\bm{k}} - \hbar\bm{k}\cdot\bm{v}_s$. As a result of this Doppler shift, the thermal conductivity in nodal superconductors is dominated by the contribution from delocalized quasiparticles, leading to an initial steep increase of $\kappa(H)/T\propto \sqrt{H}$ for line nodes and $\kappa(H)/T\propto \left| H\log H \right|$ for point nodes \cite{yamashita2017fully}. In contrast, in full-gap type II superconductors, all quasiparticles are bound to the vortex cores, and the magnetic field has a negligible effect on the heat transport \cite{lowell1970mixed,PhysRevB.86.064504,zhang2015nodeless} except in the vicinity of $H_{c2}$, where the vortex cores overlap.

Thermal conductivity is a directional probe sensitive to the quasiparticles with momentum parallel to the thermal current ($\bm{k}\cdot \bm{Q}\neq 0$) and perpendicular to the magnetic field ($\bm{k}\times \bm{H} \neq 0$) because ${\bm H}\perp {\bm v_s}$. To investigate the quasiparticle excitations around the $a$ axis, we measured $\kappa$ along the $a$ axis (${\bm Q}||a$) for ${\bm H}||c$ and ${\bm H}||a$ [Figs.\,3(b) and 3(c)]. As seen in the insets, while the former geometry sensitively probes the point node at the $a$ axis, the latter geometry selectively probes quasiparticle excitations from point nodes that are off the $a$ axis but have momentum in the $a$ axis direction. For the very-clean crystal (\#1), $\kappa/T$ collapses into a single curve and decreases almost linearly below $\sim$0.25\,K for both ${\bm H}||c$ and ${\bm H}||a$. For the ultra-clean crystal (\#2), $\kappa/T$ normalized by the normal state value $\kappa_\mathrm{0N}/T$ for ${\bm H}||c$ is nearly half of that for the very-clean crystal below $\sim$ 0.3\,K [Fig.\,3(b)]. Simple linear extrapolations to $T = 0$ give negative intercepts for all data, indicating a very small residual $\kappa/T$ at $T = 0$. To obtain the zero temperature limit of $\kappa/T$ quantitatively, the data are fitted using $\kappa/T = \kappa_0/T + AT^{\alpha}$ with $\kappa_0/T \geq 0$ and $1\leq\alpha\leq2$. This yields the vanishingly small $\kappa_0/T$ at low fields for both samples (see Figs.\,S2, S3, and S4(a) in Supplemental Material \cite{supp}).

Figure\,4(a) shows $\kappa_0/T$ normalized by $\kappa_\mathrm{0N}/T$ as a function of $H/H_{c2}$. Remarkably, $\kappa_0/T$ of the very-clean crystal (\#1) is less than 1\% of the normal-state value even at $H\sim 0.09H_{c2}^c$ for ${\bm H}||c$ and at $H\sim 0.2H_{c2}^a$ for ${\bm H}||a$. Moreover, as shown in the inset, $(\kappa_0/T)/(\kappa_\mathrm{0N}/T)$ of the ultra-clean crystal (\#2) for ${\bm H}||c$ is significantly suppressed from that of the very-clean crystal. For comparison, we plot data for two other U-based superconductors, UPt$_3$ \cite{suderow1997thermal} and URu$_2$Si$_2$ \cite{PhysRevLett.99.116402}, which are believed to have both point and line nodes. In striking contrast to the present results, $\kappa_0/T$ for UPt$_3$ and URu$_2$Si$_2$ exhibits a steep increase at low fields. As shown in Fig.\,S4(b), even at 0.13\,K, $\kappa/T$ of the ultra-clean crystal is only 2\% of $\kappa_\mathrm{0N}/T$ up to $H \sim 0.12H_{c2}^c$ \cite{supp}, still much smaller than $(\kappa_0/T)/(\kappa_\mathrm{0N}/T)$ for UPt$_3$ and URu$_2$Si$_2$. These results indicate that UTe$_2$ is essentially different from other U-based unconventional superconductors in that only a very few delocalized quasiparticles are excited, even well inside the vortex state. In Fig.\,4(b), we compare $\kappa_0/T$ of UTe$_2$ with typical $s$-wave superconductor Nb \cite{lowell1970mixed} and nodal superconductor UPt$_3$ \cite{suderow1997thermal} over a wide field range. For Nb, we show the data measured in both ascending and descending magnetic fields, although the difference is small. Here we restrict the argument of $H$-dependence for ${\bm H}||c$ to the low field regime, where the normal-state magnetoresistance is small and $\kappa_\mathrm{0N}/T$ can be approximated by $L_0/\rho(H=0)$ (see Fig.\,S5 in Supplemental Material \cite{supp}). It is evident that $H$-dependence of UTe$_2$ is very close to that of Nb. These results lead us to conclude that quasiparticle excitations with the velocity component along the $a$ axis are negligibly small, ruling out the presence of point nodes at and around the $a$ axis. This indicates that $\kappa$ in zero field is dominated by the phonon contribution $\kappa_{ph}$ at low temperatures, consistent with the power law behavior $\kappa/T \propto T^{1.48}$ [see inset of Fig.\,3(a)].

The field dependence of $\kappa/T$ at 0.22\,K for ${\bm H}||c$ and ${\bm H}||a$ [Fig.\,4(c)], which are both nearly constant at low fields, further supports the absence of point nodes around the $a$ axis. It has been reported that the thermal conductivity of the $d$-wave superconductor CeCoIn$ _5$ is almost $H$-independent at very low temperatures \cite{PhysRevLett.87.057002}. This is likely because the $\sqrt{H}$ dependence of the quasiparticle density of states is canceled out by the reduction of the quasiparticle mean free path, which is proportional to intervortex distance $\propto 1/\sqrt{H}$ \cite{PhysRevB.72.214515}. However, such cancellation is not expected for point nodes. As shown in the inset of Fig.\,4(c), we find a peak structure at $\mu_0H \sim 0.8$\,T for $\bm{H}||a$. Since an anomaly in the temperature dependence of $H^a_{c2}$ below 1\,T has been reported \cite{PhysRevResearch.2.032014}, the observed peak may be related to this anomaly. Although the origin of this peak should be scrutinized, this does not influence the analysis of the residual term $\kappa_0/T$.

\section{Discussion}
Having established the absence of any type of nodes at and around the $a$ axis, we discuss the superconducting order parameter belonging to the irreducible representation of the $D_{2h}$ point group in UTe$ _2$. Our results rule out the $B_{3u}$ state with point nodes along the $a$ axis. NMR measurements reported a clear reduction of the Knight shift for $\bm{H}||b$ and $\bm{H}||c$ \cite{PhysRevB.103.L100503,fujibayashi2022superconducting}. These rule out the possibilities of the $B_{1u}$ and $B_{2u}$ states. It should be noted that since the Knight shift measurements for $\bm{H}||a$ contain large error bars, it is premature to conclude the $B_{3u}$ state pointed out in Ref.\,\citenum{fujibayashi2022superconducting}. Based on these results, we conclude that the superconducting order parameter in UTe$ _2$ is represented by the $A_u$ symmetry.

Here we comment on the results of the specific heat. The quadratic temperature dependence of $C/T$ has been reported in the very clean crystal with RRR=220 \cite{aoki2022first}, suggesting the presence of point nodes. However, we point out that the heat capacity data of UTe$_2$ need to be scrutinized because they include contributions from both local and itinerant excitations and may include local excitations that are not directly related to the quasiparticles excited from the superconducting condensate. Indeed, recent $\mathrm{\mu}$SR experiments have reported that uranium defects induce local magnetic clusters that give a finite magnetic contribution to $C/T$ \cite{sundar2023ubiquitous}. Given this situation in UTe$_2$, thermal conductivity, which is not influenced by such local excitations, is a more direct bulk probe for determining the superconducting gap structure of this material.

In addition to the spin-triplet full-gap superconductivity, an unexpected and unique feature of the superconducting state of UTe$_2$ is that the quasiparticle excitations are extremely sensitive to disorder due to impurities/defects. In Fig.\,4(a) and Fig.\,S6 \cite{supp}, we compare $(\kappa_0/T)/(\kappa_\mathrm{0N}/T)$ for crystals with different RRR as a function of $H/H_{c2}$ and RRR, respectively. The data of RRR = 22 are taken from Ref.\,\citenum{PhysRevB.100.220504}. Surprisingly, as shown in the inset of Fig.\,4(a), $(\kappa_0/T)/(\kappa_\mathrm{0N}/T)$ of the ultra-clean crystal (\#2) is one order of magnitude smaller than that of the very-clean crystal (\#1) at $H/H_{c2} \sim 0.09$, indicating that the quasiparticle excitations is still strongly affected by the disorder even in the ultra-clean region where RRR is well above 260. It is well-known that impurities have a considerable effect on quasiparticle excitations in unconventional superconductors, and this appears to be more pronounced in UTe$_2$. The extreme sensitivity of quasiparticle excitations to the disorder should provide important clues to the mechanism of superconductivity in UTe$_2$ which still remains elusive. It has been pointed out that the field dependence of $\kappa/T$ of the moderately clean crystal bears a resemblance to that of superconductors with point nodes \cite{PhysRevB.100.220504}. Thus, the present results, combined with the previous results \cite{PhysRevB.100.220504}, indicate that the superconducting gap narrows in some directions with increasing disorder, leading to the formation of gap anisotropy that mimics point nodes.

A possible explanation for the extreme sensitivity is that a $B_{3u}$ dominant state is induced by disorder, which gives rise to the quasiparticles with momentum along the $a$ axis. A plausible origin of the disorder is uranium defects whose concentration is very sensitive to sample growth conditions \cite{PhysRevMaterials.6.073401}. It has been suggested that the uranium defects locally disrupt long-range magnetic correlations, and cause magnetic clusters to form \cite{sundar2023ubiquitous,tokunaga2022slow,iguchi2022microscopic}. For moderately clean crystals, the suppression of magnetic fluctuations may induce the $B_{3u}$ dominant phase. In fact, recent magnetocaloric experiments \cite{tokiwa2022stabilization} have reported that a field-induced phase transition to the $B_{3u}$ dominant phase may be caused by a metamagnetic crossover accompanied by a change in magnetic fluctuations.

The field applied parallel to the $a$ axis may induce the pair potential of $B_{3u}$ symmetry to that of $A_u$ symmetry. Indeed, the substantial mixing of $B_{3u}$ symmetry with $A_u$ symmetry induces the emergence of point nodes around the $a$ axis as illustrated in the inset of Fig.\,3(c). In the present very-clean crystal, vanishingly small $(\kappa_0/T)/(\kappa_\mathrm{0N}/T)$ for $\bm{Q}||\bm{H}||a$ precludes the possible emergence of such point nodes at least up to $H\sim 0.2 H^{a}_{c2}$. Although a more detailed investigation is needed, the observed steep increase in $\kappa_0/T$ above 6T may indicate a transition to the $B_{3u}$ dominant phase \cite{tokiwa2022stabilization}. 

We emphasize that the fully gapped $A_u$ symmetry is the first realization of spin-triplet superconductivity analogous to the B phase of superfluid $^3$He. Recent theoretical studies have proposed that the topological properties of this phase depend on the shape of the Fermi surface \cite{PhysRevLett.123.217001}. It should be noted that UTe$_2$ likely has the 3D Fermi surface. In fact, recent measurements of angle-resolved photoemission spectroscopy reveal the presence of 3D electronic structure with closed Fermi surface along the $c$ axis \cite{PhysRevLett.124.076401,fujimori2019electronic}. The 3D structure is further supported by the almost isotropic normal state resistivity \cite{PhysRevB.106.L060505} and the nearly identical coherence lengths along the $a$ and $c$ axes, $\xi_a \approx \xi_c$ [Fig.\,1(e)]. Although very recent quantum oscillation measurements reported the presence of 2D Fermi surfaces \cite{aoki2022first}, this does not exclude the 3D Fermi surface. This is because the topology of the Fermi surfaces may change in strong magnetic fields due to a Lifshitz transition \cite{PhysRevLett.124.086601}. Moreover, the electronic specific heat coefficient $\gamma_e \sim 100$ mJ/K$ ^2$mol obtained from the quantum oscillations is still smaller than $\gamma_e \sim$ 120 mJ/K$ ^2$mol from the specific heat, implying the existence of undiscovered Fermi surface. Given the 3D Fermi surface, the $A_u$ state is characterized by a nontrivial 3D winding number \cite{PhysRevLett.123.217001}. In this case, UTe$ _2$ is a long-sought 3D strong topological superconductor with emergent Majorana surface states on any crystal plane.

\begin{acknowledgments}
We thank K. Ishihara, T. Shibauchi, and Y. Tokunaga for insightful discussions. This work is supported by Grants-in-Aid for Scientific Research (KAKENHI) (Nos. 18H05227, 18H03680, 18H01180, 21K13881) and on Innovative Areas ``Quantum Liquid Crystals'' (No. 19H05824) from the Japan Society for the Promotion of Science, and JST CREST (JPMJCR19T5).
\end{acknowledgments}


\begin{thebibliography}{53}%
	\makeatletter
	\providecommand \@ifxundefined [1]{%
	 \@ifx{#1\undefined}
	}%
	\providecommand \@ifnum [1]{%
	 \ifnum #1\expandafter \@firstoftwo
	 \else \expandafter \@secondoftwo
	 \fi
	}%
	\providecommand \@ifx [1]{%
	 \ifx #1\expandafter \@firstoftwo
	 \else \expandafter \@secondoftwo
	 \fi
	}%
	\providecommand \natexlab [1]{#1}%
	\providecommand \enquote  [1]{``#1''}%
	\providecommand \bibnamefont  [1]{#1}%
	\providecommand \bibfnamefont [1]{#1}%
	\providecommand \citenamefont [1]{#1}%
	\providecommand \href@noop [0]{\@secondoftwo}%
	\providecommand \href [0]{\begingroup \@sanitize@url \@href}%
	\providecommand \@href[1]{\@@startlink{#1}\@@href}%
	\providecommand \@@href[1]{\endgroup#1\@@endlink}%
	\providecommand \@sanitize@url [0]{\catcode `\\12\catcode `\$12\catcode
	  `\&12\catcode `\#12\catcode `\^12\catcode `\_12\catcode `\%12\relax}%
	\providecommand \@@startlink[1]{}%
	\providecommand \@@endlink[0]{}%
	\providecommand \url  [0]{\begingroup\@sanitize@url \@url }%
	\providecommand \@url [1]{\endgroup\@href {#1}{\urlprefix }}%
	\providecommand \urlprefix  [0]{URL }%
	\providecommand \Eprint [0]{\href }%
	\providecommand \doibase [0]{https://doi.org/}%
	\providecommand \selectlanguage [0]{\@gobble}%
	\providecommand \bibinfo  [0]{\@secondoftwo}%
	\providecommand \bibfield  [0]{\@secondoftwo}%
	\providecommand \translation [1]{[#1]}%
	\providecommand \BibitemOpen [0]{}%
	\providecommand \bibitemStop [0]{}%
	\providecommand \bibitemNoStop [0]{.\EOS\space}%
	\providecommand \EOS [0]{\spacefactor3000\relax}%
	\providecommand \BibitemShut  [1]{\csname bibitem#1\endcsname}%
	\let\auto@bib@innerbib\@empty
	\bibitem [{\citenamefont {Sato}\ and\ \citenamefont
	  {Ando}(2017)}]{sato2017topological}%
	  \BibitemOpen
	  \bibfield  {author} {\bibinfo {author} {\bibfnamefont {M.}~\bibnamefont
	  {Sato}}\ and\ \bibinfo {author} {\bibfnamefont {Y.}~\bibnamefont {Ando}},\
	  }\bibfield  {title} {\bibinfo {title} {Topological superconductors: a
	  review},\ }\href@noop {} {\bibfield  {journal} {\bibinfo  {journal} {Rep.
	  Prog. Phys.}\ }\textbf {\bibinfo {volume} {80}},\ \bibinfo {pages} {076501}
	  (\bibinfo {year} {2017})}\BibitemShut {NoStop}%
	\bibitem [{\citenamefont {Ivanov}(2001)}]{PhysRevLett.86.268}%
	  \BibitemOpen
	  \bibfield  {author} {\bibinfo {author} {\bibfnamefont {D.~A.}\ \bibnamefont
	  {Ivanov}},\ }\bibfield  {title} {\bibinfo {title} {{Non-Abelian Statistics of
	  Half-Quantum Vortices in $\mathit{p}$-Wave Superconductors}},\ }\href
	  {https://doi.org/10.1103/PhysRevLett.86.268} {\bibfield  {journal} {\bibinfo
	  {journal} {Phys. Rev. Lett.}\ }\textbf {\bibinfo {volume} {86}},\ \bibinfo
	  {pages} {268} (\bibinfo {year} {2001})}\BibitemShut {NoStop}%
	\bibitem [{\citenamefont {Kitaev}(2003)}]{Kitaev2003}%
	  \BibitemOpen
	  \bibfield  {author} {\bibinfo {author} {\bibfnamefont {A.~Y.}\ \bibnamefont
	  {Kitaev}},\ }\bibfield  {title} {\bibinfo {title} {Fault-tolerant quantum
	  computation by anyons},\ }\href@noop {} {\bibfield  {journal} {\bibinfo
	  {journal} {Ann. Phys.}\ }\textbf {\bibinfo {volume} {303}},\ \bibinfo {pages}
	  {2} (\bibinfo {year} {2003})}\BibitemShut {NoStop}%
	\bibitem [{\citenamefont {Leggett}(1975)}]{RevModPhys.47.331}%
	  \BibitemOpen
	  \bibfield  {author} {\bibinfo {author} {\bibfnamefont {A.~J.}\ \bibnamefont
	  {Leggett}},\ }\bibfield  {title} {\bibinfo {title} {{A theoretical
	  description of the new phases of liquid $^{3}\mathrm{He}$}},\ }\href
	  {https://doi.org/10.1103/RevModPhys.47.331} {\bibfield  {journal} {\bibinfo
	  {journal} {Rev. Mod. Phys.}\ }\textbf {\bibinfo {volume} {47}},\ \bibinfo
	  {pages} {331} (\bibinfo {year} {1975})}\BibitemShut {NoStop}%
	\bibitem [{\citenamefont {Ran}\ \emph {et~al.}(2019{\natexlab{a}})\citenamefont
	  {Ran}, \citenamefont {Eckberg}, \citenamefont {Ding}, \citenamefont
	  {Furukawa}, \citenamefont {Metz}, \citenamefont {Saha}, \citenamefont {Liu},
	  \citenamefont {Zic}, \citenamefont {Kim}, \citenamefont {Paglione} \emph
	  {et~al.}}]{ran2019nearly}%
	  \BibitemOpen
	  \bibfield  {author} {\bibinfo {author} {\bibfnamefont {S.}~\bibnamefont
	  {Ran}}, \bibinfo {author} {\bibfnamefont {C.}~\bibnamefont {Eckberg}},
	  \bibinfo {author} {\bibfnamefont {Q.-P.}\ \bibnamefont {Ding}}, \bibinfo
	  {author} {\bibfnamefont {Y.}~\bibnamefont {Furukawa}}, \bibinfo {author}
	  {\bibfnamefont {T.}~\bibnamefont {Metz}}, \bibinfo {author} {\bibfnamefont
	  {S.~R.}\ \bibnamefont {Saha}}, \bibinfo {author} {\bibfnamefont {I.-L.}\
	  \bibnamefont {Liu}}, \bibinfo {author} {\bibfnamefont {M.}~\bibnamefont
	  {Zic}}, \bibinfo {author} {\bibfnamefont {H.}~\bibnamefont {Kim}}, \bibinfo
	  {author} {\bibfnamefont {J.}~\bibnamefont {Paglione}}, \emph {et~al.},\
	  }\bibfield  {title} {\bibinfo {title} {Nearly ferromagnetic spin-triplet
	  superconductivity},\ }\href@noop {} {\bibfield  {journal} {\bibinfo
	  {journal} {Science}\ }\textbf {\bibinfo {volume} {365}},\ \bibinfo {pages}
	  {684} (\bibinfo {year} {2019}{\natexlab{a}})}\BibitemShut {NoStop}%
	\bibitem [{\citenamefont {Aoki}\ \emph {et~al.}(2019)\citenamefont {Aoki},
	  \citenamefont {Nakamura}, \citenamefont {Honda}, \citenamefont {Li},
	  \citenamefont {Homma}, \citenamefont {Shimizu}, \citenamefont {Sato},
	  \citenamefont {Knebel}, \citenamefont {Brison}, \citenamefont {Pourret} \emph
	  {et~al.}}]{aoki2019unconventional}%
	  \BibitemOpen
	  \bibfield  {author} {\bibinfo {author} {\bibfnamefont {D.}~\bibnamefont
	  {Aoki}}, \bibinfo {author} {\bibfnamefont {A.}~\bibnamefont {Nakamura}},
	  \bibinfo {author} {\bibfnamefont {F.}~\bibnamefont {Honda}}, \bibinfo
	  {author} {\bibfnamefont {D.}~\bibnamefont {Li}}, \bibinfo {author}
	  {\bibfnamefont {Y.}~\bibnamefont {Homma}}, \bibinfo {author} {\bibfnamefont
	  {Y.}~\bibnamefont {Shimizu}}, \bibinfo {author} {\bibfnamefont {Y.~J.}\
	  \bibnamefont {Sato}}, \bibinfo {author} {\bibfnamefont {G.}~\bibnamefont
	  {Knebel}}, \bibinfo {author} {\bibfnamefont {J.-P.}\ \bibnamefont {Brison}},
	  \bibinfo {author} {\bibfnamefont {A.}~\bibnamefont {Pourret}}, \emph
	  {et~al.},\ }\bibfield  {title} {\bibinfo {title} {{Unconventional
	  superconductivity in heavy fermion UTe$ _2$}},\ }\href@noop {} {\bibfield
	  {journal} {\bibinfo  {journal} {J. Phys. Soc. Jpn.}\ }\textbf {\bibinfo
	  {volume} {88}},\ \bibinfo {pages} {043702} (\bibinfo {year}
	  {2019})}\BibitemShut {NoStop}%
	\bibitem [{\citenamefont {Ran}\ \emph {et~al.}(2019{\natexlab{b}})\citenamefont
	  {Ran}, \citenamefont {Liu}, \citenamefont {Eo}, \citenamefont {Campbell},
	  \citenamefont {Neves}, \citenamefont {Fuhrman}, \citenamefont {Saha},
	  \citenamefont {Eckberg}, \citenamefont {Kim}, \citenamefont {Graf} \emph
	  {et~al.}}]{ran2019extreme}%
	  \BibitemOpen
	  \bibfield  {author} {\bibinfo {author} {\bibfnamefont {S.}~\bibnamefont
	  {Ran}}, \bibinfo {author} {\bibfnamefont {I.-L.}\ \bibnamefont {Liu}},
	  \bibinfo {author} {\bibfnamefont {Y.~S.}\ \bibnamefont {Eo}}, \bibinfo
	  {author} {\bibfnamefont {D.~J.}\ \bibnamefont {Campbell}}, \bibinfo {author}
	  {\bibfnamefont {P.~M.}\ \bibnamefont {Neves}}, \bibinfo {author}
	  {\bibfnamefont {W.~T.}\ \bibnamefont {Fuhrman}}, \bibinfo {author}
	  {\bibfnamefont {S.~R.}\ \bibnamefont {Saha}}, \bibinfo {author}
	  {\bibfnamefont {C.}~\bibnamefont {Eckberg}}, \bibinfo {author} {\bibfnamefont
	  {H.}~\bibnamefont {Kim}}, \bibinfo {author} {\bibfnamefont {D.}~\bibnamefont
	  {Graf}}, \emph {et~al.},\ }\bibfield  {title} {\bibinfo {title} {Extreme
	  magnetic field-boosted superconductivity},\ }\href@noop {} {\bibfield
	  {journal} {\bibinfo  {journal} {Nat. Phys.}\ }\textbf {\bibinfo {volume}
	  {15}},\ \bibinfo {pages} {1250} (\bibinfo {year}
	  {2019}{\natexlab{b}})}\BibitemShut {NoStop}%
	\bibitem [{\citenamefont {Knafo}\ \emph {et~al.}(2021)\citenamefont {Knafo},
	  \citenamefont {Nardone}, \citenamefont {Vali{\v{s}}ka}, \citenamefont
	  {Zitouni}, \citenamefont {Lapertot}, \citenamefont {Aoki}, \citenamefont
	  {Knebel},\ and\ \citenamefont {Braithwaite}}]{knafo2021comparison}%
	  \BibitemOpen
	  \bibfield  {author} {\bibinfo {author} {\bibfnamefont {W.}~\bibnamefont
	  {Knafo}}, \bibinfo {author} {\bibfnamefont {M.}~\bibnamefont {Nardone}},
	  \bibinfo {author} {\bibfnamefont {M.}~\bibnamefont {Vali{\v{s}}ka}}, \bibinfo
	  {author} {\bibfnamefont {A.}~\bibnamefont {Zitouni}}, \bibinfo {author}
	  {\bibfnamefont {G.}~\bibnamefont {Lapertot}}, \bibinfo {author}
	  {\bibfnamefont {D.}~\bibnamefont {Aoki}}, \bibinfo {author} {\bibfnamefont
	  {G.}~\bibnamefont {Knebel}},\ and\ \bibinfo {author} {\bibfnamefont
	  {D.}~\bibnamefont {Braithwaite}},\ }\bibfield  {title} {\bibinfo {title}
	  {{Comparison of two superconducting phases induced by a magnetic field in
	  UTe$ _2$}},\ }\href@noop {} {\bibfield  {journal} {\bibinfo  {journal}
	  {Commun. Phys.}\ }\textbf {\bibinfo {volume} {4}},\ \bibinfo {pages} {40}
	  (\bibinfo {year} {2021})}\BibitemShut {NoStop}%
	\bibitem [{\citenamefont {Knebel}\ \emph {et~al.}(2019)\citenamefont {Knebel},
	  \citenamefont {Knafo}, \citenamefont {Pourret}, \citenamefont {Niu},
	  \citenamefont {Vali{\v{s}}ka}, \citenamefont {Braithwaite}, \citenamefont
	  {Lapertot}, \citenamefont {Nardone}, \citenamefont {Zitouni}, \citenamefont
	  {Mishra} \emph {et~al.}}]{knebel2019field}%
	  \BibitemOpen
	  \bibfield  {author} {\bibinfo {author} {\bibfnamefont {G.}~\bibnamefont
	  {Knebel}}, \bibinfo {author} {\bibfnamefont {W.}~\bibnamefont {Knafo}},
	  \bibinfo {author} {\bibfnamefont {A.}~\bibnamefont {Pourret}}, \bibinfo
	  {author} {\bibfnamefont {Q.}~\bibnamefont {Niu}}, \bibinfo {author}
	  {\bibfnamefont {M.}~\bibnamefont {Vali{\v{s}}ka}}, \bibinfo {author}
	  {\bibfnamefont {D.}~\bibnamefont {Braithwaite}}, \bibinfo {author}
	  {\bibfnamefont {G.}~\bibnamefont {Lapertot}}, \bibinfo {author}
	  {\bibfnamefont {M.}~\bibnamefont {Nardone}}, \bibinfo {author} {\bibfnamefont
	  {A.}~\bibnamefont {Zitouni}}, \bibinfo {author} {\bibfnamefont
	  {S.}~\bibnamefont {Mishra}}, \emph {et~al.},\ }\bibfield  {title} {\bibinfo
	  {title} {{Field-reentrant superconductivity close to a metamagnetic
	  transition in the heavy-fermion superconductor UTe$ _2$}},\ }\href@noop {}
	  {\bibfield  {journal} {\bibinfo  {journal} {J. Phys. Soc. Jpn.}\ }\textbf
	  {\bibinfo {volume} {88}},\ \bibinfo {pages} {063707} (\bibinfo {year}
	  {2019})}\BibitemShut {NoStop}%
	\bibitem [{\citenamefont {Nakamine}\ \emph {et~al.}(2019)\citenamefont
	  {Nakamine}, \citenamefont {Kitagawa}, \citenamefont {Ishida}, \citenamefont
	  {Tokunaga}, \citenamefont {Sakai}, \citenamefont {Kambe}, \citenamefont
	  {Nakamura}, \citenamefont {Shimizu}, \citenamefont {Homma}, \citenamefont
	  {Li} \emph {et~al.}}]{nakamine2019superconducting}%
	  \BibitemOpen
	  \bibfield  {author} {\bibinfo {author} {\bibfnamefont {G.}~\bibnamefont
	  {Nakamine}}, \bibinfo {author} {\bibfnamefont {S.}~\bibnamefont {Kitagawa}},
	  \bibinfo {author} {\bibfnamefont {K.}~\bibnamefont {Ishida}}, \bibinfo
	  {author} {\bibfnamefont {Y.}~\bibnamefont {Tokunaga}}, \bibinfo {author}
	  {\bibfnamefont {H.}~\bibnamefont {Sakai}}, \bibinfo {author} {\bibfnamefont
	  {S.}~\bibnamefont {Kambe}}, \bibinfo {author} {\bibfnamefont
	  {A.}~\bibnamefont {Nakamura}}, \bibinfo {author} {\bibfnamefont
	  {Y.}~\bibnamefont {Shimizu}}, \bibinfo {author} {\bibfnamefont
	  {Y.}~\bibnamefont {Homma}}, \bibinfo {author} {\bibfnamefont
	  {D.}~\bibnamefont {Li}}, \emph {et~al.},\ }\bibfield  {title} {\bibinfo
	  {title} {{Superconducting properties of heavy fermion UTe$_2$ revealed by
	  $^{125}$Te-nuclear magnetic resonance}},\ }\href@noop {} {\bibfield
	  {journal} {\bibinfo  {journal} {J. Phys. Soc. Jpn.}\ }\textbf {\bibinfo
	  {volume} {88}},\ \bibinfo {pages} {113703} (\bibinfo {year}
	  {2019})}\BibitemShut {NoStop}%
	\bibitem [{\citenamefont {Nakamine}\ \emph {et~al.}(2021)\citenamefont
	  {Nakamine}, \citenamefont {Kinjo}, \citenamefont {Kitagawa}, \citenamefont
	  {Ishida}, \citenamefont {Tokunaga}, \citenamefont {Sakai}, \citenamefont
	  {Kambe}, \citenamefont {Nakamura}, \citenamefont {Shimizu}, \citenamefont
	  {Homma}, \citenamefont {Li}, \citenamefont {Honda},\ and\ \citenamefont
	  {Aoki}}]{PhysRevB.103.L100503}%
	  \BibitemOpen
	  \bibfield  {author} {\bibinfo {author} {\bibfnamefont {G.}~\bibnamefont
	  {Nakamine}}, \bibinfo {author} {\bibfnamefont {K.}~\bibnamefont {Kinjo}},
	  \bibinfo {author} {\bibfnamefont {S.}~\bibnamefont {Kitagawa}}, \bibinfo
	  {author} {\bibfnamefont {K.}~\bibnamefont {Ishida}}, \bibinfo {author}
	  {\bibfnamefont {Y.}~\bibnamefont {Tokunaga}}, \bibinfo {author}
	  {\bibfnamefont {H.}~\bibnamefont {Sakai}}, \bibinfo {author} {\bibfnamefont
	  {S.}~\bibnamefont {Kambe}}, \bibinfo {author} {\bibfnamefont
	  {A.}~\bibnamefont {Nakamura}}, \bibinfo {author} {\bibfnamefont
	  {Y.}~\bibnamefont {Shimizu}}, \bibinfo {author} {\bibfnamefont
	  {Y.}~\bibnamefont {Homma}}, \bibinfo {author} {\bibfnamefont
	  {D.}~\bibnamefont {Li}}, \bibinfo {author} {\bibfnamefont {F.}~\bibnamefont
	  {Honda}},\ and\ \bibinfo {author} {\bibfnamefont {D.}~\bibnamefont {Aoki}},\
	  }\bibfield  {title} {\bibinfo {title} {{Anisotropic response of spin
	  susceptibility in the superconducting state of ${\mathrm{UTe}}_{2}$ probed
	  with $^{125}\mathrm{Te}\text{\ensuremath{-}}\mathrm{NMR}$ measurement}},\
	  }\href {https://doi.org/10.1103/PhysRevB.103.L100503} {\bibfield  {journal}
	  {\bibinfo  {journal} {Phys. Rev. B}\ }\textbf {\bibinfo {volume} {103}},\
	  \bibinfo {pages} {L100503} (\bibinfo {year} {2021})}\BibitemShut {NoStop}%
	\bibitem [{\citenamefont {Fujibayashi}\ \emph {et~al.}(2022)\citenamefont
	  {Fujibayashi}, \citenamefont {Nakamine}, \citenamefont {Kinjo}, \citenamefont
	  {Kitagawa}, \citenamefont {Ishida}, \citenamefont {Tokunaga}, \citenamefont
	  {Sakai}, \citenamefont {Kambe}, \citenamefont {Nakamura}, \citenamefont
	  {Shimizu} \emph {et~al.}}]{fujibayashi2022superconducting}%
	  \BibitemOpen
	  \bibfield  {author} {\bibinfo {author} {\bibfnamefont {H.}~\bibnamefont
	  {Fujibayashi}}, \bibinfo {author} {\bibfnamefont {G.}~\bibnamefont
	  {Nakamine}}, \bibinfo {author} {\bibfnamefont {K.}~\bibnamefont {Kinjo}},
	  \bibinfo {author} {\bibfnamefont {S.}~\bibnamefont {Kitagawa}}, \bibinfo
	  {author} {\bibfnamefont {K.}~\bibnamefont {Ishida}}, \bibinfo {author}
	  {\bibfnamefont {Y.}~\bibnamefont {Tokunaga}}, \bibinfo {author}
	  {\bibfnamefont {H.}~\bibnamefont {Sakai}}, \bibinfo {author} {\bibfnamefont
	  {S.}~\bibnamefont {Kambe}}, \bibinfo {author} {\bibfnamefont
	  {A.}~\bibnamefont {Nakamura}}, \bibinfo {author} {\bibfnamefont
	  {Y.}~\bibnamefont {Shimizu}}, \emph {et~al.},\ }\bibfield  {title} {\bibinfo
	  {title} {{Superconducting Order Parameter in UTe$ _2$ Determined by Knight
	  Shift Measurement}},\ }\href@noop {} {\bibfield  {journal} {\bibinfo
	  {journal} {J. Phys. Soc. Jpn.}\ }\textbf {\bibinfo {volume} {91}},\ \bibinfo
	  {pages} {043705} (\bibinfo {year} {2022})}\BibitemShut {NoStop}%
	\bibitem [{\citenamefont {Saxena}\ \emph {et~al.}(2000)\citenamefont {Saxena},
	  \citenamefont {Agarwal}, \citenamefont {Ahilan}, \citenamefont {Grosche},
	  \citenamefont {Haselwimmer}, \citenamefont {Steiner}, \citenamefont {Pugh},
	  \citenamefont {Walker}, \citenamefont {Julian}, \citenamefont {Monthoux}
	  \emph {et~al.}}]{saxena2000superconductivity}%
	  \BibitemOpen
	  \bibfield  {author} {\bibinfo {author} {\bibfnamefont {S.}~\bibnamefont
	  {Saxena}}, \bibinfo {author} {\bibfnamefont {P.}~\bibnamefont {Agarwal}},
	  \bibinfo {author} {\bibfnamefont {K.}~\bibnamefont {Ahilan}}, \bibinfo
	  {author} {\bibfnamefont {F.}~\bibnamefont {Grosche}}, \bibinfo {author}
	  {\bibfnamefont {R.}~\bibnamefont {Haselwimmer}}, \bibinfo {author}
	  {\bibfnamefont {M.}~\bibnamefont {Steiner}}, \bibinfo {author} {\bibfnamefont
	  {E.}~\bibnamefont {Pugh}}, \bibinfo {author} {\bibfnamefont {I.}~\bibnamefont
	  {Walker}}, \bibinfo {author} {\bibfnamefont {S.}~\bibnamefont {Julian}},
	  \bibinfo {author} {\bibfnamefont {P.}~\bibnamefont {Monthoux}}, \emph
	  {et~al.},\ }\bibfield  {title} {\bibinfo {title} {{Superconductivity on the
	  border of itinerant-electron ferromagnetism in UGe$ _2$}},\ }\href@noop {}
	  {\bibfield  {journal} {\bibinfo  {journal} {Nature}\ }\textbf {\bibinfo
	  {volume} {406}},\ \bibinfo {pages} {587} (\bibinfo {year}
	  {2000})}\BibitemShut {NoStop}%
	\bibitem [{\citenamefont {Aoki}\ \emph {et~al.}(2001)\citenamefont {Aoki},
	  \citenamefont {Huxley}, \citenamefont {Ressouche}, \citenamefont
	  {Braithwaite}, \citenamefont {Flouquet}, \citenamefont {Brison},
	  \citenamefont {Lhotel},\ and\ \citenamefont {Paulsen}}]{aoki2001coexistence}%
	  \BibitemOpen
	  \bibfield  {author} {\bibinfo {author} {\bibfnamefont {D.}~\bibnamefont
	  {Aoki}}, \bibinfo {author} {\bibfnamefont {A.}~\bibnamefont {Huxley}},
	  \bibinfo {author} {\bibfnamefont {E.}~\bibnamefont {Ressouche}}, \bibinfo
	  {author} {\bibfnamefont {D.}~\bibnamefont {Braithwaite}}, \bibinfo {author}
	  {\bibfnamefont {J.}~\bibnamefont {Flouquet}}, \bibinfo {author}
	  {\bibfnamefont {J.-P.}\ \bibnamefont {Brison}}, \bibinfo {author}
	  {\bibfnamefont {E.}~\bibnamefont {Lhotel}},\ and\ \bibinfo {author}
	  {\bibfnamefont {C.}~\bibnamefont {Paulsen}},\ }\bibfield  {title} {\bibinfo
	  {title} {{Coexistence of superconductivity and ferromagnetism in URhGe}},\
	  }\href@noop {} {\bibfield  {journal} {\bibinfo  {journal} {Nature}\ }\textbf
	  {\bibinfo {volume} {413}},\ \bibinfo {pages} {613} (\bibinfo {year}
	  {2001})}\BibitemShut {NoStop}%
	\bibitem [{\citenamefont {Huy}\ \emph {et~al.}(2007)\citenamefont {Huy},
	  \citenamefont {Gasparini}, \citenamefont {de~Nijs}, \citenamefont {Huang},
	  \citenamefont {Klaasse}, \citenamefont {Gortenmulder}, \citenamefont
	  {de~Visser}, \citenamefont {Hamann}, \citenamefont {G\"orlach},\ and\
	  \citenamefont {L\"ohneysen}}]{PhysRevLett.99.067006}%
	  \BibitemOpen
	  \bibfield  {author} {\bibinfo {author} {\bibfnamefont {N.~T.}\ \bibnamefont
	  {Huy}}, \bibinfo {author} {\bibfnamefont {A.}~\bibnamefont {Gasparini}},
	  \bibinfo {author} {\bibfnamefont {D.~E.}\ \bibnamefont {de~Nijs}}, \bibinfo
	  {author} {\bibfnamefont {Y.}~\bibnamefont {Huang}}, \bibinfo {author}
	  {\bibfnamefont {J.~C.~P.}\ \bibnamefont {Klaasse}}, \bibinfo {author}
	  {\bibfnamefont {T.}~\bibnamefont {Gortenmulder}}, \bibinfo {author}
	  {\bibfnamefont {A.}~\bibnamefont {de~Visser}}, \bibinfo {author}
	  {\bibfnamefont {A.}~\bibnamefont {Hamann}}, \bibinfo {author} {\bibfnamefont
	  {T.}~\bibnamefont {G\"orlach}},\ and\ \bibinfo {author} {\bibfnamefont
	  {H.~v.}\ \bibnamefont {L\"ohneysen}},\ }\bibfield  {title} {\bibinfo {title}
	  {{Superconductivity on the Border of Weak Itinerant Ferromagnetism in
	  UCoGe}},\ }\href {https://doi.org/10.1103/PhysRevLett.99.067006} {\bibfield
	  {journal} {\bibinfo  {journal} {Phys. Rev. Lett.}\ }\textbf {\bibinfo
	  {volume} {99}},\ \bibinfo {pages} {067006} (\bibinfo {year}
	  {2007})}\BibitemShut {NoStop}%
	\bibitem [{\citenamefont {Hayes}\ \emph {et~al.}(2021)\citenamefont {Hayes},
	  \citenamefont {Wei}, \citenamefont {Metz}, \citenamefont {Zhang},
	  \citenamefont {Eo}, \citenamefont {Ran}, \citenamefont {Saha}, \citenamefont
	  {Collini}, \citenamefont {Butch}, \citenamefont {Agterberg} \emph
	  {et~al.}}]{hayes2021multicomponent}%
	  \BibitemOpen
	  \bibfield  {author} {\bibinfo {author} {\bibfnamefont {I.}~\bibnamefont
	  {Hayes}}, \bibinfo {author} {\bibfnamefont {D.~S.}\ \bibnamefont {Wei}},
	  \bibinfo {author} {\bibfnamefont {T.}~\bibnamefont {Metz}}, \bibinfo {author}
	  {\bibfnamefont {J.}~\bibnamefont {Zhang}}, \bibinfo {author} {\bibfnamefont
	  {Y.}~\bibnamefont {Eo}}, \bibinfo {author} {\bibfnamefont {S.}~\bibnamefont
	  {Ran}}, \bibinfo {author} {\bibfnamefont {S.}~\bibnamefont {Saha}}, \bibinfo
	  {author} {\bibfnamefont {J.}~\bibnamefont {Collini}}, \bibinfo {author}
	  {\bibfnamefont {N.}~\bibnamefont {Butch}}, \bibinfo {author} {\bibfnamefont
	  {D.}~\bibnamefont {Agterberg}}, \emph {et~al.},\ }\bibfield  {title}
	  {\bibinfo {title} {{Multicomponent superconducting order parameter in UTe$
	  _2$}},\ }\href@noop {} {\bibfield  {journal} {\bibinfo  {journal} {Science}\
	  }\textbf {\bibinfo {volume} {373}},\ \bibinfo {pages} {797} (\bibinfo {year}
	  {2021})}\BibitemShut {NoStop}%
	\bibitem [{\citenamefont {Jiao}\ \emph {et~al.}(2020)\citenamefont {Jiao},
	  \citenamefont {Howard}, \citenamefont {Ran}, \citenamefont {Wang},
	  \citenamefont {Rodriguez}, \citenamefont {Sigrist}, \citenamefont {Wang},
	  \citenamefont {Butch},\ and\ \citenamefont {Madhavan}}]{jiao2020chiral}%
	  \BibitemOpen
	  \bibfield  {author} {\bibinfo {author} {\bibfnamefont {L.}~\bibnamefont
	  {Jiao}}, \bibinfo {author} {\bibfnamefont {S.}~\bibnamefont {Howard}},
	  \bibinfo {author} {\bibfnamefont {S.}~\bibnamefont {Ran}}, \bibinfo {author}
	  {\bibfnamefont {Z.}~\bibnamefont {Wang}}, \bibinfo {author} {\bibfnamefont
	  {J.~O.}\ \bibnamefont {Rodriguez}}, \bibinfo {author} {\bibfnamefont
	  {M.}~\bibnamefont {Sigrist}}, \bibinfo {author} {\bibfnamefont
	  {Z.}~\bibnamefont {Wang}}, \bibinfo {author} {\bibfnamefont {N.~P.}\
	  \bibnamefont {Butch}},\ and\ \bibinfo {author} {\bibfnamefont
	  {V.}~\bibnamefont {Madhavan}},\ }\bibfield  {title} {\bibinfo {title}
	  {{Chiral superconductivity in heavy-fermion metal UTe$ _2$}},\ }\href@noop {}
	  {\bibfield  {journal} {\bibinfo  {journal} {Nature}\ }\textbf {\bibinfo
	  {volume} {579}},\ \bibinfo {pages} {523} (\bibinfo {year}
	  {2020})}\BibitemShut {NoStop}%
	\bibitem [{\citenamefont {Cairns}\ \emph {et~al.}(2020)\citenamefont {Cairns},
	  \citenamefont {Stevens}, \citenamefont {D~O’Neill},\ and\ \citenamefont
	  {Huxley}}]{cairns2020composition}%
	  \BibitemOpen
	  \bibfield  {author} {\bibinfo {author} {\bibfnamefont {L.~P.}\ \bibnamefont
	  {Cairns}}, \bibinfo {author} {\bibfnamefont {C.~R.}\ \bibnamefont {Stevens}},
	  \bibinfo {author} {\bibfnamefont {C.}~\bibnamefont {D~O’Neill}},\ and\
	  \bibinfo {author} {\bibfnamefont {A.}~\bibnamefont {Huxley}},\ }\bibfield
	  {title} {\bibinfo {title} {{Composition dependence of the superconducting
	  properties of UTe$ _2$}},\ }\href@noop {} {\bibfield  {journal} {\bibinfo
	  {journal} {J. Phys. Condens. Matter.}\ }\textbf {\bibinfo {volume} {32}},\
	  \bibinfo {pages} {415602} (\bibinfo {year} {2020})}\BibitemShut {NoStop}%
	\bibitem [{\citenamefont {Thomas}\ \emph {et~al.}(2021)\citenamefont {Thomas},
	  \citenamefont {Stevens}, \citenamefont {Santos}, \citenamefont {Fender},
	  \citenamefont {Bauer}, \citenamefont {Ronning}, \citenamefont {Thompson},
	  \citenamefont {Huxley},\ and\ \citenamefont {Rosa}}]{PhysRevB.104.224501}%
	  \BibitemOpen
	  \bibfield  {author} {\bibinfo {author} {\bibfnamefont {S.~M.}\ \bibnamefont
	  {Thomas}}, \bibinfo {author} {\bibfnamefont {C.}~\bibnamefont {Stevens}},
	  \bibinfo {author} {\bibfnamefont {F.~B.}\ \bibnamefont {Santos}}, \bibinfo
	  {author} {\bibfnamefont {S.~S.}\ \bibnamefont {Fender}}, \bibinfo {author}
	  {\bibfnamefont {E.~D.}\ \bibnamefont {Bauer}}, \bibinfo {author}
	  {\bibfnamefont {F.}~\bibnamefont {Ronning}}, \bibinfo {author} {\bibfnamefont
	  {J.~D.}\ \bibnamefont {Thompson}}, \bibinfo {author} {\bibfnamefont
	  {A.}~\bibnamefont {Huxley}},\ and\ \bibinfo {author} {\bibfnamefont
	  {P.~F.~S.}\ \bibnamefont {Rosa}},\ }\bibfield  {title} {\bibinfo {title}
	  {{Spatially inhomogeneous superconductivity in ${\mathrm{UTe}}_{2}$}},\
	  }\href {https://doi.org/10.1103/PhysRevB.104.224501} {\bibfield  {journal}
	  {\bibinfo  {journal} {Phys. Rev. B}\ }\textbf {\bibinfo {volume} {104}},\
	  \bibinfo {pages} {224501} (\bibinfo {year} {2021})}\BibitemShut {NoStop}%
	\bibitem [{\citenamefont {Rosa}\ \emph {et~al.}(2022)\citenamefont {Rosa},
	  \citenamefont {Weiland}, \citenamefont {Fender}, \citenamefont {Scott},
	  \citenamefont {Ronning}, \citenamefont {Thompson}, \citenamefont {Bauer},\
	  and\ \citenamefont {Thomas}}]{rosa2022single}%
	  \BibitemOpen
	  \bibfield  {author} {\bibinfo {author} {\bibfnamefont {P.~F.}\ \bibnamefont
	  {Rosa}}, \bibinfo {author} {\bibfnamefont {A.}~\bibnamefont {Weiland}},
	  \bibinfo {author} {\bibfnamefont {S.~S.}\ \bibnamefont {Fender}}, \bibinfo
	  {author} {\bibfnamefont {B.~L.}\ \bibnamefont {Scott}}, \bibinfo {author}
	  {\bibfnamefont {F.}~\bibnamefont {Ronning}}, \bibinfo {author} {\bibfnamefont
	  {J.~D.}\ \bibnamefont {Thompson}}, \bibinfo {author} {\bibfnamefont {E.~D.}\
	  \bibnamefont {Bauer}},\ and\ \bibinfo {author} {\bibfnamefont {S.~M.}\
	  \bibnamefont {Thomas}},\ }\bibfield  {title} {\bibinfo {title} {{Single
	  thermodynamic transition at 2 K in superconducting UTe$ _2$ single
	  crystals}},\ }\href@noop {} {\bibfield  {journal} {\bibinfo  {journal}
	  {Commun. Mater.}\ }\textbf {\bibinfo {volume} {3}},\ \bibinfo {pages} {33}
	  (\bibinfo {year} {2022})}\BibitemShut {NoStop}%
	\bibitem [{\citenamefont {Girod}\ \emph {et~al.}(2022)\citenamefont {Girod},
	  \citenamefont {Stevens}, \citenamefont {Huxley}, \citenamefont {Bauer},
	  \citenamefont {Santos}, \citenamefont {Thompson}, \citenamefont {Fernandes},
	  \citenamefont {Zhu}, \citenamefont {Ronning}, \citenamefont {Rosa},\ and\
	  \citenamefont {Thomas}}]{girod2022thermodynamic}%
	  \BibitemOpen
	  \bibfield  {author} {\bibinfo {author} {\bibfnamefont {C.}~\bibnamefont
	  {Girod}}, \bibinfo {author} {\bibfnamefont {C.~R.}\ \bibnamefont {Stevens}},
	  \bibinfo {author} {\bibfnamefont {A.}~\bibnamefont {Huxley}}, \bibinfo
	  {author} {\bibfnamefont {E.~D.}\ \bibnamefont {Bauer}}, \bibinfo {author}
	  {\bibfnamefont {F.~B.}\ \bibnamefont {Santos}}, \bibinfo {author}
	  {\bibfnamefont {J.~D.}\ \bibnamefont {Thompson}}, \bibinfo {author}
	  {\bibfnamefont {R.~M.}\ \bibnamefont {Fernandes}}, \bibinfo {author}
	  {\bibfnamefont {J.-X.}\ \bibnamefont {Zhu}}, \bibinfo {author} {\bibfnamefont
	  {F.}~\bibnamefont {Ronning}}, \bibinfo {author} {\bibfnamefont {P.~F.~S.}\
	  \bibnamefont {Rosa}},\ and\ \bibinfo {author} {\bibfnamefont {S.~M.}\
	  \bibnamefont {Thomas}},\ }\bibfield  {title} {\bibinfo {title}
	  {{Thermodynamic and electrical transport properties of ${\mathrm{UTe}}_{2}$
	  under uniaxial stress}},\ }\href
	  {https://doi.org/10.1103/PhysRevB.106.L121101} {\bibfield  {journal}
	  {\bibinfo  {journal} {Phys. Rev. B}\ }\textbf {\bibinfo {volume} {106}},\
	  \bibinfo {pages} {L121101} (\bibinfo {year} {2022})}\BibitemShut {NoStop}%
	\bibitem [{\citenamefont {Ajeesh}\ \emph {et~al.}(2023)\citenamefont {Ajeesh},
	  \citenamefont {Bordelon}, \citenamefont {Girod}, \citenamefont {Mishra},
	  \citenamefont {Ronning}, \citenamefont {Bauer}, \citenamefont {Maiorov},
	  \citenamefont {Thompson}, \citenamefont {Rosa},\ and\ \citenamefont
	  {Thomas}}]{ajeesh2023fate}%
	  \BibitemOpen
	  \bibfield  {author} {\bibinfo {author} {\bibfnamefont {M.}~\bibnamefont
	  {Ajeesh}}, \bibinfo {author} {\bibfnamefont {M.}~\bibnamefont {Bordelon}},
	  \bibinfo {author} {\bibfnamefont {C.}~\bibnamefont {Girod}}, \bibinfo
	  {author} {\bibfnamefont {S.}~\bibnamefont {Mishra}}, \bibinfo {author}
	  {\bibfnamefont {F.}~\bibnamefont {Ronning}}, \bibinfo {author} {\bibfnamefont
	  {E.}~\bibnamefont {Bauer}}, \bibinfo {author} {\bibfnamefont
	  {B.}~\bibnamefont {Maiorov}}, \bibinfo {author} {\bibfnamefont
	  {J.}~\bibnamefont {Thompson}}, \bibinfo {author} {\bibfnamefont
	  {P.}~\bibnamefont {Rosa}},\ and\ \bibinfo {author} {\bibfnamefont
	  {S.}~\bibnamefont {Thomas}},\ }\bibfield  {title} {\bibinfo {title} {{The
	  fate of time-reversal symmetry breaking in UTe$_2$}},\ }\href@noop {}
	  {\bibfield  {journal} {\bibinfo  {journal} {arXiv preprint arXiv:2305.00589}\
	  } (\bibinfo {year} {2023})}\BibitemShut {NoStop}%
	\bibitem [{\citenamefont {Aoki}\ \emph
	  {et~al.}(2022{\natexlab{a}})\citenamefont {Aoki}, \citenamefont {Brison},
	  \citenamefont {Flouquet}, \citenamefont {Ishida}, \citenamefont {Knebel},
	  \citenamefont {Tokunaga},\ and\ \citenamefont
	  {Yanase}}]{aoki2022unconventional}%
	  \BibitemOpen
	  \bibfield  {author} {\bibinfo {author} {\bibfnamefont {D.}~\bibnamefont
	  {Aoki}}, \bibinfo {author} {\bibfnamefont {J.-P.}\ \bibnamefont {Brison}},
	  \bibinfo {author} {\bibfnamefont {J.}~\bibnamefont {Flouquet}}, \bibinfo
	  {author} {\bibfnamefont {K.}~\bibnamefont {Ishida}}, \bibinfo {author}
	  {\bibfnamefont {G.}~\bibnamefont {Knebel}}, \bibinfo {author} {\bibfnamefont
	  {Y.}~\bibnamefont {Tokunaga}},\ and\ \bibinfo {author} {\bibfnamefont
	  {Y.}~\bibnamefont {Yanase}},\ }\bibfield  {title} {\bibinfo {title}
	  {{Unconventional superconductivity in UTe$ _2$}},\ }\href@noop {} {\bibfield
	  {journal} {\bibinfo  {journal} {J. Phys. Condens. Matter.}\ }\textbf
	  {\bibinfo {volume} {34}},\ \bibinfo {pages} {243002} (\bibinfo {year}
	  {2022}{\natexlab{a}})}\BibitemShut {NoStop}%
	\bibitem [{\citenamefont {Metz}\ \emph {et~al.}(2019)\citenamefont {Metz},
	  \citenamefont {Bae}, \citenamefont {Ran}, \citenamefont {Liu}, \citenamefont
	  {Eo}, \citenamefont {Fuhrman}, \citenamefont {Agterberg}, \citenamefont
	  {Anlage}, \citenamefont {Butch},\ and\ \citenamefont
	  {Paglione}}]{PhysRevB.100.220504}%
	  \BibitemOpen
	  \bibfield  {author} {\bibinfo {author} {\bibfnamefont {T.}~\bibnamefont
	  {Metz}}, \bibinfo {author} {\bibfnamefont {S.}~\bibnamefont {Bae}}, \bibinfo
	  {author} {\bibfnamefont {S.}~\bibnamefont {Ran}}, \bibinfo {author}
	  {\bibfnamefont {I.-L.}\ \bibnamefont {Liu}}, \bibinfo {author} {\bibfnamefont
	  {Y.~S.}\ \bibnamefont {Eo}}, \bibinfo {author} {\bibfnamefont {W.~T.}\
	  \bibnamefont {Fuhrman}}, \bibinfo {author} {\bibfnamefont {D.~F.}\
	  \bibnamefont {Agterberg}}, \bibinfo {author} {\bibfnamefont {S.~M.}\
	  \bibnamefont {Anlage}}, \bibinfo {author} {\bibfnamefont {N.~P.}\
	  \bibnamefont {Butch}},\ and\ \bibinfo {author} {\bibfnamefont
	  {J.}~\bibnamefont {Paglione}},\ }\bibfield  {title} {\bibinfo {title}
	  {{Point-node gap structure of the spin-triplet superconductor
	  ${\mathrm{UTe}}_{2}$}},\ }\href {https://doi.org/10.1103/PhysRevB.100.220504}
	  {\bibfield  {journal} {\bibinfo  {journal} {Phys. Rev. B}\ }\textbf {\bibinfo
	  {volume} {100}},\ \bibinfo {pages} {220504(R)} (\bibinfo {year}
	  {2019})}\BibitemShut {NoStop}%
	\bibitem [{\citenamefont {Bae}\ \emph {et~al.}(2021)\citenamefont {Bae},
	  \citenamefont {Kim}, \citenamefont {Eo}, \citenamefont {Ran}, \citenamefont
	  {Liu}, \citenamefont {Fuhrman}, \citenamefont {Paglione}, \citenamefont
	  {Butch}, \citenamefont {Anlage} \emph {et~al.}}]{bae2021anomalous}%
	  \BibitemOpen
	  \bibfield  {author} {\bibinfo {author} {\bibfnamefont {S.}~\bibnamefont
	  {Bae}}, \bibinfo {author} {\bibfnamefont {H.}~\bibnamefont {Kim}}, \bibinfo
	  {author} {\bibfnamefont {Y.~S.}\ \bibnamefont {Eo}}, \bibinfo {author}
	  {\bibfnamefont {S.}~\bibnamefont {Ran}}, \bibinfo {author} {\bibfnamefont
	  {I.-l.}\ \bibnamefont {Liu}}, \bibinfo {author} {\bibfnamefont {W.~T.}\
	  \bibnamefont {Fuhrman}}, \bibinfo {author} {\bibfnamefont {J.}~\bibnamefont
	  {Paglione}}, \bibinfo {author} {\bibfnamefont {N.~P.}\ \bibnamefont {Butch}},
	  \bibinfo {author} {\bibfnamefont {S.~M.}\ \bibnamefont {Anlage}}, \emph
	  {et~al.},\ }\bibfield  {title} {\bibinfo {title} {{Anomalous normal fluid
	  response in a chiral superconductor UTe$ _2$}},\ }\href@noop {} {\bibfield
	  {journal} {\bibinfo  {journal} {Nat. Commun.}\ }\textbf {\bibinfo {volume}
	  {12}},\ \bibinfo {pages} {1} (\bibinfo {year} {2021})}\BibitemShut {NoStop}%
	\bibitem [{\citenamefont {Ishihara}\ \emph {et~al.}(2021)\citenamefont
	  {Ishihara}, \citenamefont {Roppongi}, \citenamefont {Kobayashi},
	  \citenamefont {Mizukami}, \citenamefont {Sakai}, \citenamefont {Haga},
	  \citenamefont {Hashimoto},\ and\ \citenamefont
	  {Shibauchi}}]{ishihara2021chiral}%
	  \BibitemOpen
	  \bibfield  {author} {\bibinfo {author} {\bibfnamefont {K.}~\bibnamefont
	  {Ishihara}}, \bibinfo {author} {\bibfnamefont {M.}~\bibnamefont {Roppongi}},
	  \bibinfo {author} {\bibfnamefont {M.}~\bibnamefont {Kobayashi}}, \bibinfo
	  {author} {\bibfnamefont {Y.}~\bibnamefont {Mizukami}}, \bibinfo {author}
	  {\bibfnamefont {H.}~\bibnamefont {Sakai}}, \bibinfo {author} {\bibfnamefont
	  {Y.}~\bibnamefont {Haga}}, \bibinfo {author} {\bibfnamefont {K.}~\bibnamefont
	  {Hashimoto}},\ and\ \bibinfo {author} {\bibfnamefont {T.}~\bibnamefont
	  {Shibauchi}},\ }\bibfield  {title} {\bibinfo {title} {{Chiral
	  superconductivity in UTe$ _2$ probed by anisotropic low-energy
	  excitations}},\ }\href@noop {} {\bibfield  {journal} {\bibinfo  {journal}
	  {arXiv preprint arXiv:2105.13721}\ } (\bibinfo {year} {2021})}\BibitemShut
	  {NoStop}%
	\bibitem [{\citenamefont {Kittaka}\ \emph {et~al.}(2020)\citenamefont
	  {Kittaka}, \citenamefont {Shimizu}, \citenamefont {Sakakibara}, \citenamefont
	  {Nakamura}, \citenamefont {Li}, \citenamefont {Homma}, \citenamefont {Honda},
	  \citenamefont {Aoki},\ and\ \citenamefont
	  {Machida}}]{PhysRevResearch.2.032014}%
	  \BibitemOpen
	  \bibfield  {author} {\bibinfo {author} {\bibfnamefont {S.}~\bibnamefont
	  {Kittaka}}, \bibinfo {author} {\bibfnamefont {Y.}~\bibnamefont {Shimizu}},
	  \bibinfo {author} {\bibfnamefont {T.}~\bibnamefont {Sakakibara}}, \bibinfo
	  {author} {\bibfnamefont {A.}~\bibnamefont {Nakamura}}, \bibinfo {author}
	  {\bibfnamefont {D.}~\bibnamefont {Li}}, \bibinfo {author} {\bibfnamefont
	  {Y.}~\bibnamefont {Homma}}, \bibinfo {author} {\bibfnamefont
	  {F.}~\bibnamefont {Honda}}, \bibinfo {author} {\bibfnamefont
	  {D.}~\bibnamefont {Aoki}},\ and\ \bibinfo {author} {\bibfnamefont
	  {K.}~\bibnamefont {Machida}},\ }\bibfield  {title} {\bibinfo {title}
	  {{Orientation of point nodes and nonunitary triplet pairing tuned by the
	  easy-axis magnetization in ${\mathrm{UTe}}_{2}$}},\ }\href
	  {https://doi.org/10.1103/PhysRevResearch.2.032014} {\bibfield  {journal}
	  {\bibinfo  {journal} {Phys. Rev. Res.}\ }\textbf {\bibinfo {volume} {2}},\
	  \bibinfo {pages} {032014(R)} (\bibinfo {year} {2020})}\BibitemShut {NoStop}%
	\bibitem [{\citenamefont {Mineev}(2022)}]{mineev2022low}%
	  \BibitemOpen
	  \bibfield  {author} {\bibinfo {author} {\bibfnamefont {V.~P.}\ \bibnamefont
	  {Mineev}},\ }\bibfield  {title} {\bibinfo {title} {{Low Temperature Specific
	  Heat and Thermal Conductivity in Superconducting UTe$ _2$}},\ }\href@noop {}
	  {\bibfield  {journal} {\bibinfo  {journal} {J. Phys. Soc. Jpn.}\ }\textbf
	  {\bibinfo {volume} {91}},\ \bibinfo {pages} {074601} (\bibinfo {year}
	  {2022})}\BibitemShut {NoStop}%
	\bibitem [{\citenamefont {Ishizuka}\ \emph {et~al.}(2019)\citenamefont
	  {Ishizuka}, \citenamefont {Sumita}, \citenamefont {Daido},\ and\
	  \citenamefont {Yanase}}]{PhysRevLett.123.217001}%
	  \BibitemOpen
	  \bibfield  {author} {\bibinfo {author} {\bibfnamefont {J.}~\bibnamefont
	  {Ishizuka}}, \bibinfo {author} {\bibfnamefont {S.}~\bibnamefont {Sumita}},
	  \bibinfo {author} {\bibfnamefont {A.}~\bibnamefont {Daido}},\ and\ \bibinfo
	  {author} {\bibfnamefont {Y.}~\bibnamefont {Yanase}},\ }\bibfield  {title}
	  {\bibinfo {title} {{Insulator-Metal Transition and Topological
	  Superconductivity in ${\mathrm{UTe}}_{2}$ from a First-Principles
	  Calculation}},\ }\href {https://doi.org/10.1103/PhysRevLett.123.217001}
	  {\bibfield  {journal} {\bibinfo  {journal} {Phys. Rev. Lett.}\ }\textbf
	  {\bibinfo {volume} {123}},\ \bibinfo {pages} {217001} (\bibinfo {year}
	  {2019})}\BibitemShut {NoStop}%
	\bibitem [{\citenamefont {Aoki}\ \emph
	  {et~al.}(2022{\natexlab{b}})\citenamefont {Aoki}, \citenamefont {Sakai},
	  \citenamefont {Opletal}, \citenamefont {Tokiwa}, \citenamefont {Ishizuka},
	  \citenamefont {Yanase}, \citenamefont {Harima}, \citenamefont {Nakamura},
	  \citenamefont {Li}, \citenamefont {Homma} \emph {et~al.}}]{aoki2022first}%
	  \BibitemOpen
	  \bibfield  {author} {\bibinfo {author} {\bibfnamefont {D.}~\bibnamefont
	  {Aoki}}, \bibinfo {author} {\bibfnamefont {H.}~\bibnamefont {Sakai}},
	  \bibinfo {author} {\bibfnamefont {P.}~\bibnamefont {Opletal}}, \bibinfo
	  {author} {\bibfnamefont {Y.}~\bibnamefont {Tokiwa}}, \bibinfo {author}
	  {\bibfnamefont {J.}~\bibnamefont {Ishizuka}}, \bibinfo {author}
	  {\bibfnamefont {Y.}~\bibnamefont {Yanase}}, \bibinfo {author} {\bibfnamefont
	  {H.}~\bibnamefont {Harima}}, \bibinfo {author} {\bibfnamefont
	  {A.}~\bibnamefont {Nakamura}}, \bibinfo {author} {\bibfnamefont
	  {D.}~\bibnamefont {Li}}, \bibinfo {author} {\bibfnamefont {Y.}~\bibnamefont
	  {Homma}}, \emph {et~al.},\ }\bibfield  {title} {\bibinfo {title} {{First
	  Observation of the de Haas--van Alphen Effect and Fermi Surfaces in the
	  Unconventional Superconductor UTe$ _2$}},\ }\href@noop {} {\bibfield
	  {journal} {\bibinfo  {journal} {J. Phys. Soc. Jpn.}\ }\textbf {\bibinfo
	  {volume} {91}},\ \bibinfo {pages} {083704} (\bibinfo {year}
	  {2022}{\natexlab{b}})}\BibitemShut {NoStop}%
	\bibitem [{\citenamefont {Sakai}\ \emph {et~al.}(2022)\citenamefont {Sakai},
	  \citenamefont {Opletal}, \citenamefont {Tokiwa}, \citenamefont {Yamamoto},
	  \citenamefont {Tokunaga}, \citenamefont {Kambe},\ and\ \citenamefont
	  {Haga}}]{PhysRevMaterials.6.073401}%
	  \BibitemOpen
	  \bibfield  {author} {\bibinfo {author} {\bibfnamefont {H.}~\bibnamefont
	  {Sakai}}, \bibinfo {author} {\bibfnamefont {P.}~\bibnamefont {Opletal}},
	  \bibinfo {author} {\bibfnamefont {Y.}~\bibnamefont {Tokiwa}}, \bibinfo
	  {author} {\bibfnamefont {E.}~\bibnamefont {Yamamoto}}, \bibinfo {author}
	  {\bibfnamefont {Y.}~\bibnamefont {Tokunaga}}, \bibinfo {author}
	  {\bibfnamefont {S.}~\bibnamefont {Kambe}},\ and\ \bibinfo {author}
	  {\bibfnamefont {Y.}~\bibnamefont {Haga}},\ }\bibfield  {title} {\bibinfo
	  {title} {{Single crystal growth of superconducting ${\mathrm{UTe}}_{2}$ by
	  molten salt flux method}},\ }\href
	  {https://doi.org/10.1103/PhysRevMaterials.6.073401} {\bibfield  {journal}
	  {\bibinfo  {journal} {Phys. Rev. Mater.}\ }\textbf {\bibinfo {volume} {6}},\
	  \bibinfo {pages} {073401} (\bibinfo {year} {2022})}\BibitemShut {NoStop}%
	\bibitem [{\citenamefont {Matsuda}\ \emph {et~al.}(2006)\citenamefont
	  {Matsuda}, \citenamefont {Izawa},\ and\ \citenamefont
	  {Vekhter}}]{matsuda2006nodal}%
	  \BibitemOpen
	  \bibfield  {author} {\bibinfo {author} {\bibfnamefont {Y.}~\bibnamefont
	  {Matsuda}}, \bibinfo {author} {\bibfnamefont {K.}~\bibnamefont {Izawa}},\
	  and\ \bibinfo {author} {\bibfnamefont {I.}~\bibnamefont {Vekhter}},\
	  }\bibfield  {title} {\bibinfo {title} {Nodal structure of unconventional
	  superconductors probed by angle resolved thermal transport measurements},\
	  }\href@noop {} {\bibfield  {journal} {\bibinfo  {journal} {J. Phys. Condens.
	  Matter.}\ }\textbf {\bibinfo {volume} {18}},\ \bibinfo {pages} {R705}
	  (\bibinfo {year} {2006})}\BibitemShut {NoStop}%
	\bibitem [{\citenamefont {Wang}\ \emph {et~al.}(2001)\citenamefont {Wang},
	  \citenamefont {Revaz}, \citenamefont {Erb},\ and\ \citenamefont
	  {Junod}}]{PhysRevB.63.094508}%
	  \BibitemOpen
	  \bibfield  {author} {\bibinfo {author} {\bibfnamefont {Y.}~\bibnamefont
	  {Wang}}, \bibinfo {author} {\bibfnamefont {B.}~\bibnamefont {Revaz}},
	  \bibinfo {author} {\bibfnamefont {A.}~\bibnamefont {Erb}},\ and\ \bibinfo
	  {author} {\bibfnamefont {A.}~\bibnamefont {Junod}},\ }\bibfield  {title}
	  {\bibinfo {title} {{Direct observation and anisotropy of the contribution of
	  gap nodes in the low-temperature specific heat of
	  ${\mathrm{YBa}}_{2}{\mathrm{Cu}}_{3}{\mathrm{O}}_{7}$}},\ }\href
	  {https://doi.org/10.1103/PhysRevB.63.094508} {\bibfield  {journal} {\bibinfo
	  {journal} {Phys. Rev. B}\ }\textbf {\bibinfo {volume} {63}},\ \bibinfo
	  {pages} {094508} (\bibinfo {year} {2001})}\BibitemShut {NoStop}%
	\bibitem [{\citenamefont {Taylor}\ \emph {et~al.}(2007)\citenamefont {Taylor},
	  \citenamefont {Carrington},\ and\ \citenamefont
	  {Schlueter}}]{PhysRevLett.99.057001}%
	  \BibitemOpen
	  \bibfield  {author} {\bibinfo {author} {\bibfnamefont {O.~J.}\ \bibnamefont
	  {Taylor}}, \bibinfo {author} {\bibfnamefont {A.}~\bibnamefont {Carrington}},\
	  and\ \bibinfo {author} {\bibfnamefont {J.~A.}\ \bibnamefont {Schlueter}},\
	  }\bibfield  {title} {\bibinfo {title} {{Specific-Heat Measurements of the Gap
	  Structure of the Organic Superconductors
	  $\ensuremath{\kappa}\mathrm{\text{\ensuremath{-}}}(\mathrm{ET}{)}_{2}\mathrm{Cu}[\mathrm{N}(\mathrm{CN}{)}_{2}]\mathrm{Br}$
	  and
	  $\ensuremath{\kappa}\mathrm{\text{\ensuremath{-}}}(\mathrm{ET}{)}_{2}\mathrm{Cu}(\mathrm{NCS}{)}_{2}$}},\
	  }\href {https://doi.org/10.1103/PhysRevLett.99.057001} {\bibfield  {journal}
	  {\bibinfo  {journal} {Phys. Rev. Lett.}\ }\textbf {\bibinfo {volume} {99}},\
	  \bibinfo {pages} {057001} (\bibinfo {year} {2007})}\BibitemShut {NoStop}%
	\bibitem[{\citenamefont{supp}(2023)}]{supp}
	  \bibinfo{author}{\bibinfo{note}{See Supplemental Material at [URL will be inserted by publisher] for additional data.}}
	
	\bibitem [{\citenamefont {Kadowaki}\ and\ \citenamefont
	  {Woods}(1986)}]{kadowaki1986universal}%
	  \BibitemOpen
	  \bibfield  {author} {\bibinfo {author} {\bibfnamefont {K.}~\bibnamefont
	  {Kadowaki}}\ and\ \bibinfo {author} {\bibfnamefont {S.}~\bibnamefont
	  {Woods}},\ }\bibfield  {title} {\bibinfo {title} {Universal relationship of
	  the resistivity and specific heat in heavy-fermion compounds},\ }\href@noop
	  {} {\bibfield  {journal} {\bibinfo  {journal} {Solid State Commun.}\ }\textbf
	  {\bibinfo {volume} {58}},\ \bibinfo {pages} {507} (\bibinfo {year}
	  {1986})}\BibitemShut {NoStop}%
	\bibitem [{\citenamefont {Yu}\ \emph {et~al.}(1992)\citenamefont {Yu},
	  \citenamefont {Salamon}, \citenamefont {Lu},\ and\ \citenamefont
	  {Lee}}]{PhysRevLett.69.1431}%
	  \BibitemOpen
	  \bibfield  {author} {\bibinfo {author} {\bibfnamefont {R.~C.}\ \bibnamefont
	  {Yu}}, \bibinfo {author} {\bibfnamefont {M.~B.}\ \bibnamefont {Salamon}},
	  \bibinfo {author} {\bibfnamefont {J.~P.}\ \bibnamefont {Lu}},\ and\ \bibinfo
	  {author} {\bibfnamefont {W.~C.}\ \bibnamefont {Lee}},\ }\bibfield  {title}
	  {\bibinfo {title} {{Thermal conductivity of an untwinned
	  ${\mathrm{YBa}}_{2}$${\mathrm{Cu}}_{3}$${\mathrm{O}}_{7\mathrm{\ensuremath{-}}\mathrm{\ensuremath{\delta}}}$
	  single crystal and a new interpretation of the superconducting state thermal
	  transport}},\ }\href {https://doi.org/10.1103/PhysRevLett.69.1431} {\bibfield
	   {journal} {\bibinfo  {journal} {Phys. Rev. Lett.}\ }\textbf {\bibinfo
	  {volume} {69}},\ \bibinfo {pages} {1431} (\bibinfo {year}
	  {1992})}\BibitemShut {NoStop}%
	\bibitem [{\citenamefont {Izawa}\ \emph {et~al.}(2001)\citenamefont {Izawa},
	  \citenamefont {Yamaguchi}, \citenamefont {Matsuda}, \citenamefont {Shishido},
	  \citenamefont {Settai},\ and\ \citenamefont {Onuki}}]{PhysRevLett.87.057002}%
	  \BibitemOpen
	  \bibfield  {author} {\bibinfo {author} {\bibfnamefont {K.}~\bibnamefont
	  {Izawa}}, \bibinfo {author} {\bibfnamefont {H.}~\bibnamefont {Yamaguchi}},
	  \bibinfo {author} {\bibfnamefont {Y.}~\bibnamefont {Matsuda}}, \bibinfo
	  {author} {\bibfnamefont {H.}~\bibnamefont {Shishido}}, \bibinfo {author}
	  {\bibfnamefont {R.}~\bibnamefont {Settai}},\ and\ \bibinfo {author}
	  {\bibfnamefont {Y.}~\bibnamefont {Onuki}},\ }\bibfield  {title} {\bibinfo
	  {title} {{Angular Position of Nodes in the Superconducting Gap of Quasi-2D
	  Heavy-Fermion Superconductor ${\mathrm{CeCoIn}}_{5}$}},\ }\href
	  {https://doi.org/10.1103/PhysRevLett.87.057002} {\bibfield  {journal}
	  {\bibinfo  {journal} {Phys. Rev. Lett.}\ }\textbf {\bibinfo {volume} {87}},\
	  \bibinfo {pages} {057002} (\bibinfo {year} {2001})}\BibitemShut {NoStop}%
	\bibitem [{\citenamefont {Kasahara}\ \emph {et~al.}(2007)\citenamefont
	  {Kasahara}, \citenamefont {Iwasawa}, \citenamefont {Shishido}, \citenamefont
	  {Shibauchi}, \citenamefont {Behnia}, \citenamefont {Haga}, \citenamefont
	  {Matsuda}, \citenamefont {Onuki}, \citenamefont {Sigrist},\ and\
	  \citenamefont {Matsuda}}]{PhysRevLett.99.116402}%
	  \BibitemOpen
	  \bibfield  {author} {\bibinfo {author} {\bibfnamefont {Y.}~\bibnamefont
	  {Kasahara}}, \bibinfo {author} {\bibfnamefont {T.}~\bibnamefont {Iwasawa}},
	  \bibinfo {author} {\bibfnamefont {H.}~\bibnamefont {Shishido}}, \bibinfo
	  {author} {\bibfnamefont {T.}~\bibnamefont {Shibauchi}}, \bibinfo {author}
	  {\bibfnamefont {K.}~\bibnamefont {Behnia}}, \bibinfo {author} {\bibfnamefont
	  {Y.}~\bibnamefont {Haga}}, \bibinfo {author} {\bibfnamefont {T.~D.}\
	  \bibnamefont {Matsuda}}, \bibinfo {author} {\bibfnamefont {Y.}~\bibnamefont
	  {Onuki}}, \bibinfo {author} {\bibfnamefont {M.}~\bibnamefont {Sigrist}},\
	  and\ \bibinfo {author} {\bibfnamefont {Y.}~\bibnamefont {Matsuda}},\
	  }\bibfield  {title} {\bibinfo {title} {{Exotic Superconducting Properties in
	  the Electron-Hole-Compensated Heavy-Fermion ``Semimetal''
	  ${\mathrm{URu}}_{2}{\mathrm{Si}}_{2}$}},\ }\href
	  {https://doi.org/10.1103/PhysRevLett.99.116402} {\bibfield  {journal}
	  {\bibinfo  {journal} {Phys. Rev. Lett.}\ }\textbf {\bibinfo {volume} {99}},\
	  \bibinfo {pages} {116402} (\bibinfo {year} {2007})}\BibitemShut {NoStop}%
	\bibitem [{\citenamefont {Taillefer}\ \emph {et~al.}(1997)\citenamefont
	  {Taillefer}, \citenamefont {Lussier}, \citenamefont {Gagnon}, \citenamefont
	  {Behnia},\ and\ \citenamefont {Aubin}}]{PhysRevLett.79.483}%
	  \BibitemOpen
	  \bibfield  {author} {\bibinfo {author} {\bibfnamefont {L.}~\bibnamefont
	  {Taillefer}}, \bibinfo {author} {\bibfnamefont {B.}~\bibnamefont {Lussier}},
	  \bibinfo {author} {\bibfnamefont {R.}~\bibnamefont {Gagnon}}, \bibinfo
	  {author} {\bibfnamefont {K.}~\bibnamefont {Behnia}},\ and\ \bibinfo {author}
	  {\bibfnamefont {H.}~\bibnamefont {Aubin}},\ }\bibfield  {title} {\bibinfo
	  {title} {{Universal Heat Conduction in
	  ${\mathrm{YBa}}_{2}{\mathrm{Cu}}_{3}{\mathrm{O}}_{6.9}$}},\ }\href
	  {https://doi.org/10.1103/PhysRevLett.79.483} {\bibfield  {journal} {\bibinfo
	  {journal} {Phys. Rev. Lett.}\ }\textbf {\bibinfo {volume} {79}},\ \bibinfo
	  {pages} {483} (\bibinfo {year} {1997})}\BibitemShut {NoStop}%
	\bibitem [{\citenamefont {Yamashita}\ \emph {et~al.}(2017)\citenamefont
	  {Yamashita}, \citenamefont {Takenaka}, \citenamefont {Tokiwa}, \citenamefont
	  {Wilcox}, \citenamefont {Mizukami}, \citenamefont {Terazawa}, \citenamefont
	  {Kasahara}, \citenamefont {Kittaka}, \citenamefont {Sakakibara},
	  \citenamefont {Konczykowski} \emph {et~al.}}]{yamashita2017fully}%
	  \BibitemOpen
	  \bibfield  {author} {\bibinfo {author} {\bibfnamefont {T.}~\bibnamefont
	  {Yamashita}}, \bibinfo {author} {\bibfnamefont {T.}~\bibnamefont {Takenaka}},
	  \bibinfo {author} {\bibfnamefont {Y.}~\bibnamefont {Tokiwa}}, \bibinfo
	  {author} {\bibfnamefont {J.~A.}\ \bibnamefont {Wilcox}}, \bibinfo {author}
	  {\bibfnamefont {Y.}~\bibnamefont {Mizukami}}, \bibinfo {author}
	  {\bibfnamefont {D.}~\bibnamefont {Terazawa}}, \bibinfo {author}
	  {\bibfnamefont {Y.}~\bibnamefont {Kasahara}}, \bibinfo {author}
	  {\bibfnamefont {S.}~\bibnamefont {Kittaka}}, \bibinfo {author} {\bibfnamefont
	  {T.}~\bibnamefont {Sakakibara}}, \bibinfo {author} {\bibfnamefont
	  {M.}~\bibnamefont {Konczykowski}}, \emph {et~al.},\ }\bibfield  {title}
	  {\bibinfo {title} {{Fully gapped superconductivity with no sign change in the
	  prototypical heavy-fermion CeCu$ _2$Si$ _2$}},\ }\href@noop {} {\bibfield
	  {journal} {\bibinfo  {journal} {Sci. Adv.}\ }\textbf {\bibinfo {volume}
	  {3}},\ \bibinfo {pages} {e1601667} (\bibinfo {year} {2017})}\BibitemShut
	  {NoStop}%
	\bibitem [{\citenamefont {Lowell}\ and\ \citenamefont
	  {Sousa}(1970)}]{lowell1970mixed}%
	  \BibitemOpen
	  \bibfield  {author} {\bibinfo {author} {\bibfnamefont {J.}~\bibnamefont
	  {Lowell}}\ and\ \bibinfo {author} {\bibfnamefont {J.}~\bibnamefont {Sousa}},\
	  }\bibfield  {title} {\bibinfo {title} {{Mixed-state thermal conductivity of
	  type II superconductors}},\ }\href@noop {} {\bibfield  {journal} {\bibinfo
	  {journal} {J. Low Temp. Phys.}\ }\textbf {\bibinfo {volume} {3}},\ \bibinfo
	  {pages} {65} (\bibinfo {year} {1970})}\BibitemShut {NoStop}%
	\bibitem [{\citenamefont {Zhou}\ \emph {et~al.}(2012)\citenamefont {Zhou},
	  \citenamefont {Zhang}, \citenamefont {Hong}, \citenamefont {Pan},
	  \citenamefont {Qiu}, \citenamefont {Dong}, \citenamefont {Li},\ and\
	  \citenamefont {Li}}]{PhysRevB.86.064504}%
	  \BibitemOpen
	  \bibfield  {author} {\bibinfo {author} {\bibfnamefont {S.~Y.}\ \bibnamefont
	  {Zhou}}, \bibinfo {author} {\bibfnamefont {H.}~\bibnamefont {Zhang}},
	  \bibinfo {author} {\bibfnamefont {X.~C.}\ \bibnamefont {Hong}}, \bibinfo
	  {author} {\bibfnamefont {B.~Y.}\ \bibnamefont {Pan}}, \bibinfo {author}
	  {\bibfnamefont {X.}~\bibnamefont {Qiu}}, \bibinfo {author} {\bibfnamefont
	  {W.~N.}\ \bibnamefont {Dong}}, \bibinfo {author} {\bibfnamefont {X.~L.}\
	  \bibnamefont {Li}},\ and\ \bibinfo {author} {\bibfnamefont {S.~Y.}\
	  \bibnamefont {Li}},\ }\bibfield  {title} {\bibinfo {title} {{Nodeless
	  superconductivity in Ca${}_{3}$Ir${}_{4}$Sn${}_{13}$: Evidence from
	  quasiparticle heat transport}},\ }\href
	  {https://doi.org/10.1103/PhysRevB.86.064504} {\bibfield  {journal} {\bibinfo
	  {journal} {Phys. Rev. B}\ }\textbf {\bibinfo {volume} {86}},\ \bibinfo
	  {pages} {064504} (\bibinfo {year} {2012})}\BibitemShut {NoStop}%
	\bibitem [{\citenamefont {Zhang}\ \emph {et~al.}(2015)\citenamefont {Zhang},
	  \citenamefont {Xu}, \citenamefont {Kuo}, \citenamefont {Hong}, \citenamefont
	  {Wang}, \citenamefont {Cai}, \citenamefont {Dong}, \citenamefont {Lue},\ and\
	  \citenamefont {Li}}]{zhang2015nodeless}%
	  \BibitemOpen
	  \bibfield  {author} {\bibinfo {author} {\bibfnamefont {Z.}~\bibnamefont
	  {Zhang}}, \bibinfo {author} {\bibfnamefont {Y.}~\bibnamefont {Xu}}, \bibinfo
	  {author} {\bibfnamefont {C.}~\bibnamefont {Kuo}}, \bibinfo {author}
	  {\bibfnamefont {X.}~\bibnamefont {Hong}}, \bibinfo {author} {\bibfnamefont
	  {M.}~\bibnamefont {Wang}}, \bibinfo {author} {\bibfnamefont {P.}~\bibnamefont
	  {Cai}}, \bibinfo {author} {\bibfnamefont {J.}~\bibnamefont {Dong}}, \bibinfo
	  {author} {\bibfnamefont {C.}~\bibnamefont {Lue}},\ and\ \bibinfo {author}
	  {\bibfnamefont {S.}~\bibnamefont {Li}},\ }\bibfield  {title} {\bibinfo
	  {title} {{Nodeless superconducting gap in the caged-type superconductors Y$
	  _5$Rh$ _6$Sn$ _{18}$ and Lu$ _5$Rh$ _6$Sn$ _{18}$}},\ }\href@noop {}
	  {\bibfield  {journal} {\bibinfo  {journal} {Superconductor Science and
	  Technology}\ }\textbf {\bibinfo {volume} {28}},\ \bibinfo {pages} {105008}
	  (\bibinfo {year} {2015})}\BibitemShut {NoStop}%
	\bibitem [{\citenamefont {Suderow}\ \emph {et~al.}(1997)\citenamefont
	  {Suderow}, \citenamefont {Brison}, \citenamefont {Huxley},\ and\
	  \citenamefont {Flouquet}}]{suderow1997thermal}%
	  \BibitemOpen
	  \bibfield  {author} {\bibinfo {author} {\bibfnamefont {H.}~\bibnamefont
	  {Suderow}}, \bibinfo {author} {\bibfnamefont {J.}~\bibnamefont {Brison}},
	  \bibinfo {author} {\bibfnamefont {A.}~\bibnamefont {Huxley}},\ and\ \bibinfo
	  {author} {\bibfnamefont {J.}~\bibnamefont {Flouquet}},\ }\bibfield  {title}
	  {\bibinfo {title} {{Thermal conductivity and gap structure of the
	  superconducting phases of UPt$ _3$}},\ }\href@noop {} {\bibfield  {journal}
	  {\bibinfo  {journal} {J. Low Temp. Phys.}\ }\textbf {\bibinfo {volume}
	  {108}},\ \bibinfo {pages} {11} (\bibinfo {year} {1997})}\BibitemShut
	  {NoStop}%
	\bibitem [{\citenamefont {Kasahara}\ \emph {et~al.}(2005)\citenamefont
	  {Kasahara}, \citenamefont {Nakajima}, \citenamefont {Izawa}, \citenamefont
	  {Matsuda}, \citenamefont {Behnia}, \citenamefont {Shishido}, \citenamefont
	  {Settai},\ and\ \citenamefont {Onuki}}]{PhysRevB.72.214515}%
	  \BibitemOpen
	  \bibfield  {author} {\bibinfo {author} {\bibfnamefont {Y.}~\bibnamefont
	  {Kasahara}}, \bibinfo {author} {\bibfnamefont {Y.}~\bibnamefont {Nakajima}},
	  \bibinfo {author} {\bibfnamefont {K.}~\bibnamefont {Izawa}}, \bibinfo
	  {author} {\bibfnamefont {Y.}~\bibnamefont {Matsuda}}, \bibinfo {author}
	  {\bibfnamefont {K.}~\bibnamefont {Behnia}}, \bibinfo {author} {\bibfnamefont
	  {H.}~\bibnamefont {Shishido}}, \bibinfo {author} {\bibfnamefont
	  {R.}~\bibnamefont {Settai}},\ and\ \bibinfo {author} {\bibfnamefont
	  {Y.}~\bibnamefont {Onuki}},\ }\bibfield  {title} {\bibinfo {title}
	  {{Anomalous quasiparticle transport in the superconducting state of
	  ${\mathrm{CeCoIn}}_{5}$}},\ }\href
	  {https://doi.org/10.1103/PhysRevB.72.214515} {\bibfield  {journal} {\bibinfo
	  {journal} {Phys. Rev. B}\ }\textbf {\bibinfo {volume} {72}},\ \bibinfo
	  {pages} {214515} (\bibinfo {year} {2005})}\BibitemShut {NoStop}%
	\bibitem [{\citenamefont {Sundar}\ \emph {et~al.}(2023)\citenamefont {Sundar},
	  \citenamefont {Azari}, \citenamefont {Goeks}, \citenamefont {Gheidi},
	  \citenamefont {Abedi}, \citenamefont {Yakovlev}, \citenamefont {Dunsiger},
	  \citenamefont {Wilkinson}, \citenamefont {Blundell}, \citenamefont {Metz}
	  \emph {et~al.}}]{sundar2023ubiquitous}%
	  \BibitemOpen
	  \bibfield  {author} {\bibinfo {author} {\bibfnamefont {S.}~\bibnamefont
	  {Sundar}}, \bibinfo {author} {\bibfnamefont {N.}~\bibnamefont {Azari}},
	  \bibinfo {author} {\bibfnamefont {M.~R.}\ \bibnamefont {Goeks}}, \bibinfo
	  {author} {\bibfnamefont {S.}~\bibnamefont {Gheidi}}, \bibinfo {author}
	  {\bibfnamefont {M.}~\bibnamefont {Abedi}}, \bibinfo {author} {\bibfnamefont
	  {M.}~\bibnamefont {Yakovlev}}, \bibinfo {author} {\bibfnamefont {S.~R.}\
	  \bibnamefont {Dunsiger}}, \bibinfo {author} {\bibfnamefont {J.~M.}\
	  \bibnamefont {Wilkinson}}, \bibinfo {author} {\bibfnamefont {S.~J.}\
	  \bibnamefont {Blundell}}, \bibinfo {author} {\bibfnamefont {T.~E.}\
	  \bibnamefont {Metz}}, \emph {et~al.},\ }\bibfield  {title} {\bibinfo {title}
	  {{Ubiquitous spin freezing in the superconducting state of UTe$ _2$}},\
	  }\href@noop {} {\bibfield  {journal} {\bibinfo  {journal} {Commun. Phys.}\
	  }\textbf {\bibinfo {volume} {6}},\ \bibinfo {pages} {24} (\bibinfo {year}
	  {2023})}\BibitemShut {NoStop}%
	\bibitem [{\citenamefont {Tokunaga}\ \emph {et~al.}(2022)\citenamefont
	  {Tokunaga}, \citenamefont {Sakai}, \citenamefont {Kambe}, \citenamefont
	  {Haga}, \citenamefont {Tokiwa}, \citenamefont {Opletal}, \citenamefont
	  {Fujibayashi}, \citenamefont {Kinjo}, \citenamefont {Kitagawa}, \citenamefont
	  {Ishida} \emph {et~al.}}]{tokunaga2022slow}%
	  \BibitemOpen
	  \bibfield  {author} {\bibinfo {author} {\bibfnamefont {Y.}~\bibnamefont
	  {Tokunaga}}, \bibinfo {author} {\bibfnamefont {H.}~\bibnamefont {Sakai}},
	  \bibinfo {author} {\bibfnamefont {S.}~\bibnamefont {Kambe}}, \bibinfo
	  {author} {\bibfnamefont {Y.}~\bibnamefont {Haga}}, \bibinfo {author}
	  {\bibfnamefont {Y.}~\bibnamefont {Tokiwa}}, \bibinfo {author} {\bibfnamefont
	  {P.}~\bibnamefont {Opletal}}, \bibinfo {author} {\bibfnamefont
	  {H.}~\bibnamefont {Fujibayashi}}, \bibinfo {author} {\bibfnamefont
	  {K.}~\bibnamefont {Kinjo}}, \bibinfo {author} {\bibfnamefont
	  {S.}~\bibnamefont {Kitagawa}}, \bibinfo {author} {\bibfnamefont
	  {K.}~\bibnamefont {Ishida}}, \emph {et~al.},\ }\bibfield  {title} {\bibinfo
	  {title} {{Slow electronic dynamics in the paramagnetic state of UTe$ _2$}},\
	  }\href@noop {} {\bibfield  {journal} {\bibinfo  {journal} {J. Phys. Soc.
	  Jpn.}\ }\textbf {\bibinfo {volume} {91}},\ \bibinfo {pages} {023707}
	  (\bibinfo {year} {2022})}\BibitemShut {NoStop}%
	\bibitem [{\citenamefont {Iguchi}\ \emph {et~al.}(2022)\citenamefont {Iguchi},
	  \citenamefont {Man}, \citenamefont {Thomas}, \citenamefont {Ronning},
	  \citenamefont {Rosa},\ and\ \citenamefont {Moler}}]{iguchi2022microscopic}%
	  \BibitemOpen
	  \bibfield  {author} {\bibinfo {author} {\bibfnamefont {Y.}~\bibnamefont
	  {Iguchi}}, \bibinfo {author} {\bibfnamefont {H.}~\bibnamefont {Man}},
	  \bibinfo {author} {\bibfnamefont {S.}~\bibnamefont {Thomas}}, \bibinfo
	  {author} {\bibfnamefont {F.}~\bibnamefont {Ronning}}, \bibinfo {author}
	  {\bibfnamefont {P.~F.}\ \bibnamefont {Rosa}},\ and\ \bibinfo {author}
	  {\bibfnamefont {K.~A.}\ \bibnamefont {Moler}},\ }\bibfield  {title} {\bibinfo
	  {title} {{Microscopic imaging homogeneous and single phase superfluid density
	  in UTe$_2$}},\ }\href@noop {} {\bibfield  {journal} {\bibinfo  {journal}
	  {arXiv preprint arXiv:2210.09562}\ } (\bibinfo {year} {2022})}\BibitemShut
	  {NoStop}%
	\bibitem [{\citenamefont {Tokiwa}\ \emph {et~al.}(2022)\citenamefont {Tokiwa},
	  \citenamefont {Opletal}, \citenamefont {Sakai}, \citenamefont {Kubo},
	  \citenamefont {Yamamoto}, \citenamefont {Kambe}, \citenamefont {Kimata},
	  \citenamefont {Awaji}, \citenamefont {Sasaki}, \citenamefont {Aoki} \emph
	  {et~al.}}]{tokiwa2022stabilization}%
	  \BibitemOpen
	  \bibfield  {author} {\bibinfo {author} {\bibfnamefont {Y.}~\bibnamefont
	  {Tokiwa}}, \bibinfo {author} {\bibfnamefont {P.}~\bibnamefont {Opletal}},
	  \bibinfo {author} {\bibfnamefont {H.}~\bibnamefont {Sakai}}, \bibinfo
	  {author} {\bibfnamefont {K.}~\bibnamefont {Kubo}}, \bibinfo {author}
	  {\bibfnamefont {E.}~\bibnamefont {Yamamoto}}, \bibinfo {author}
	  {\bibfnamefont {S.}~\bibnamefont {Kambe}}, \bibinfo {author} {\bibfnamefont
	  {M.}~\bibnamefont {Kimata}}, \bibinfo {author} {\bibfnamefont
	  {S.}~\bibnamefont {Awaji}}, \bibinfo {author} {\bibfnamefont
	  {T.}~\bibnamefont {Sasaki}}, \bibinfo {author} {\bibfnamefont
	  {D.}~\bibnamefont {Aoki}}, \emph {et~al.},\ }\bibfield  {title} {\bibinfo
	  {title} {{Stabilization of superconductivity by metamagnetism in an easy-axis
	  magnetic field on UTe$ _2$}},\ }\href@noop {} {\bibfield  {journal} {\bibinfo
	   {journal} {arXiv preprint arXiv:2210.11769}\ } (\bibinfo {year}
	  {2022})}\BibitemShut {NoStop}%
	\bibitem [{\citenamefont {Miao}\ \emph {et~al.}(2020)\citenamefont {Miao},
	  \citenamefont {Liu}, \citenamefont {Xu}, \citenamefont {Kotta}, \citenamefont
	  {Kang}, \citenamefont {Ran}, \citenamefont {Paglione}, \citenamefont
	  {Kotliar}, \citenamefont {Butch}, \citenamefont {Denlinger},\ and\
	  \citenamefont {Wray}}]{PhysRevLett.124.076401}%
	  \BibitemOpen
	  \bibfield  {author} {\bibinfo {author} {\bibfnamefont {L.}~\bibnamefont
	  {Miao}}, \bibinfo {author} {\bibfnamefont {S.}~\bibnamefont {Liu}}, \bibinfo
	  {author} {\bibfnamefont {Y.}~\bibnamefont {Xu}}, \bibinfo {author}
	  {\bibfnamefont {E.~C.}\ \bibnamefont {Kotta}}, \bibinfo {author}
	  {\bibfnamefont {C.-J.}\ \bibnamefont {Kang}}, \bibinfo {author}
	  {\bibfnamefont {S.}~\bibnamefont {Ran}}, \bibinfo {author} {\bibfnamefont
	  {J.}~\bibnamefont {Paglione}}, \bibinfo {author} {\bibfnamefont
	  {G.}~\bibnamefont {Kotliar}}, \bibinfo {author} {\bibfnamefont {N.~P.}\
	  \bibnamefont {Butch}}, \bibinfo {author} {\bibfnamefont {J.~D.}\ \bibnamefont
	  {Denlinger}},\ and\ \bibinfo {author} {\bibfnamefont {L.~A.}\ \bibnamefont
	  {Wray}},\ }\bibfield  {title} {\bibinfo {title} {{Low Energy Band Structure
	  and Symmetries of ${\mathrm{UTe}}_{2}$ from Angle-Resolved Photoemission
	  Spectroscopy}},\ }\href {https://doi.org/10.1103/PhysRevLett.124.076401}
	  {\bibfield  {journal} {\bibinfo  {journal} {Phys. Rev. Lett.}\ }\textbf
	  {\bibinfo {volume} {124}},\ \bibinfo {pages} {076401} (\bibinfo {year}
	  {2020})}\BibitemShut {NoStop}%
	\bibitem [{\citenamefont {Fujimori}\ \emph {et~al.}(2019)\citenamefont
	  {Fujimori}, \citenamefont {Kawasaki}, \citenamefont {Takeda}, \citenamefont
	  {Yamagami}, \citenamefont {Nakamura}, \citenamefont {Homma},\ and\
	  \citenamefont {Aoki}}]{fujimori2019electronic}%
	  \BibitemOpen
	  \bibfield  {author} {\bibinfo {author} {\bibfnamefont {S.-i.}\ \bibnamefont
	  {Fujimori}}, \bibinfo {author} {\bibfnamefont {I.}~\bibnamefont {Kawasaki}},
	  \bibinfo {author} {\bibfnamefont {Y.}~\bibnamefont {Takeda}}, \bibinfo
	  {author} {\bibfnamefont {H.}~\bibnamefont {Yamagami}}, \bibinfo {author}
	  {\bibfnamefont {A.}~\bibnamefont {Nakamura}}, \bibinfo {author}
	  {\bibfnamefont {Y.}~\bibnamefont {Homma}},\ and\ \bibinfo {author}
	  {\bibfnamefont {D.}~\bibnamefont {Aoki}},\ }\bibfield  {title} {\bibinfo
	  {title} {{Electronic structure of UTe$ _2$ studied by photoelectron
	  spectroscopy}},\ }\href@noop {} {\bibfield  {journal} {\bibinfo  {journal}
	  {J. Phys. Soc. Jpn.}\ }\textbf {\bibinfo {volume} {88}},\ \bibinfo {pages}
	  {103701} (\bibinfo {year} {2019})}\BibitemShut {NoStop}%
	\bibitem [{\citenamefont {Eo}\ \emph {et~al.}(2022)\citenamefont {Eo},
	  \citenamefont {Liu}, \citenamefont {Saha}, \citenamefont {Kim}, \citenamefont
	  {Ran}, \citenamefont {Horn}, \citenamefont {Hodovanets}, \citenamefont
	  {Collini}, \citenamefont {Metz}, \citenamefont {Fuhrman}, \citenamefont
	  {Nevidomskyy}, \citenamefont {Denlinger}, \citenamefont {Butch},
	  \citenamefont {Fuhrer}, \citenamefont {Wray},\ and\ \citenamefont
	  {Paglione}}]{PhysRevB.106.L060505}%
	  \BibitemOpen
	  \bibfield  {author} {\bibinfo {author} {\bibfnamefont {Y.~S.}\ \bibnamefont
	  {Eo}}, \bibinfo {author} {\bibfnamefont {S.}~\bibnamefont {Liu}}, \bibinfo
	  {author} {\bibfnamefont {S.~R.}\ \bibnamefont {Saha}}, \bibinfo {author}
	  {\bibfnamefont {H.}~\bibnamefont {Kim}}, \bibinfo {author} {\bibfnamefont
	  {S.}~\bibnamefont {Ran}}, \bibinfo {author} {\bibfnamefont {J.~A.}\
	  \bibnamefont {Horn}}, \bibinfo {author} {\bibfnamefont {H.}~\bibnamefont
	  {Hodovanets}}, \bibinfo {author} {\bibfnamefont {J.}~\bibnamefont {Collini}},
	  \bibinfo {author} {\bibfnamefont {T.}~\bibnamefont {Metz}}, \bibinfo {author}
	  {\bibfnamefont {W.~T.}\ \bibnamefont {Fuhrman}}, \bibinfo {author}
	  {\bibfnamefont {A.~H.}\ \bibnamefont {Nevidomskyy}}, \bibinfo {author}
	  {\bibfnamefont {J.~D.}\ \bibnamefont {Denlinger}}, \bibinfo {author}
	  {\bibfnamefont {N.~P.}\ \bibnamefont {Butch}}, \bibinfo {author}
	  {\bibfnamefont {M.~S.}\ \bibnamefont {Fuhrer}}, \bibinfo {author}
	  {\bibfnamefont {L.~A.}\ \bibnamefont {Wray}},\ and\ \bibinfo {author}
	  {\bibfnamefont {J.}~\bibnamefont {Paglione}},\ }\bibfield  {title} {\bibinfo
	  {title} {{$c$-axis transport in ${\mathrm{UTe}}_{2}$: Evidence of
	  three-dimensional conductivity component}},\ }\href
	  {https://doi.org/10.1103/PhysRevB.106.L060505} {\bibfield  {journal}
	  {\bibinfo  {journal} {Phys. Rev. B}\ }\textbf {\bibinfo {volume} {106}},\
	  \bibinfo {pages} {L060505} (\bibinfo {year} {2022})}\BibitemShut {NoStop}%
	\bibitem [{\citenamefont {Niu}\ \emph {et~al.}(2020)\citenamefont {Niu},
	  \citenamefont {Knebel}, \citenamefont {Braithwaite}, \citenamefont {Aoki},
	  \citenamefont {Lapertot}, \citenamefont {Seyfarth}, \citenamefont {Brison},
	  \citenamefont {Flouquet},\ and\ \citenamefont
	  {Pourret}}]{PhysRevLett.124.086601}%
	  \BibitemOpen
	  \bibfield  {author} {\bibinfo {author} {\bibfnamefont {Q.}~\bibnamefont
	  {Niu}}, \bibinfo {author} {\bibfnamefont {G.}~\bibnamefont {Knebel}},
	  \bibinfo {author} {\bibfnamefont {D.}~\bibnamefont {Braithwaite}}, \bibinfo
	  {author} {\bibfnamefont {D.}~\bibnamefont {Aoki}}, \bibinfo {author}
	  {\bibfnamefont {G.}~\bibnamefont {Lapertot}}, \bibinfo {author}
	  {\bibfnamefont {G.}~\bibnamefont {Seyfarth}}, \bibinfo {author}
	  {\bibfnamefont {J.-P.}\ \bibnamefont {Brison}}, \bibinfo {author}
	  {\bibfnamefont {J.}~\bibnamefont {Flouquet}},\ and\ \bibinfo {author}
	  {\bibfnamefont {A.}~\bibnamefont {Pourret}},\ }\bibfield  {title} {\bibinfo
	  {title} {{Fermi-Surface Instability in the Heavy-Fermion Superconductor
	  ${\mathrm{UTe}}_{2}$}},\ }\href
	  {https://doi.org/10.1103/PhysRevLett.124.086601} {\bibfield  {journal}
	  {\bibinfo  {journal} {Phys. Rev. Lett.}\ }\textbf {\bibinfo {volume} {124}},\
	  \bibinfo {pages} {086601} (\bibinfo {year} {2020})}\BibitemShut {NoStop}%
\end{thebibliography}

%

\clearpage

\begin{figure}
	\includegraphics[clip,width=8cm]{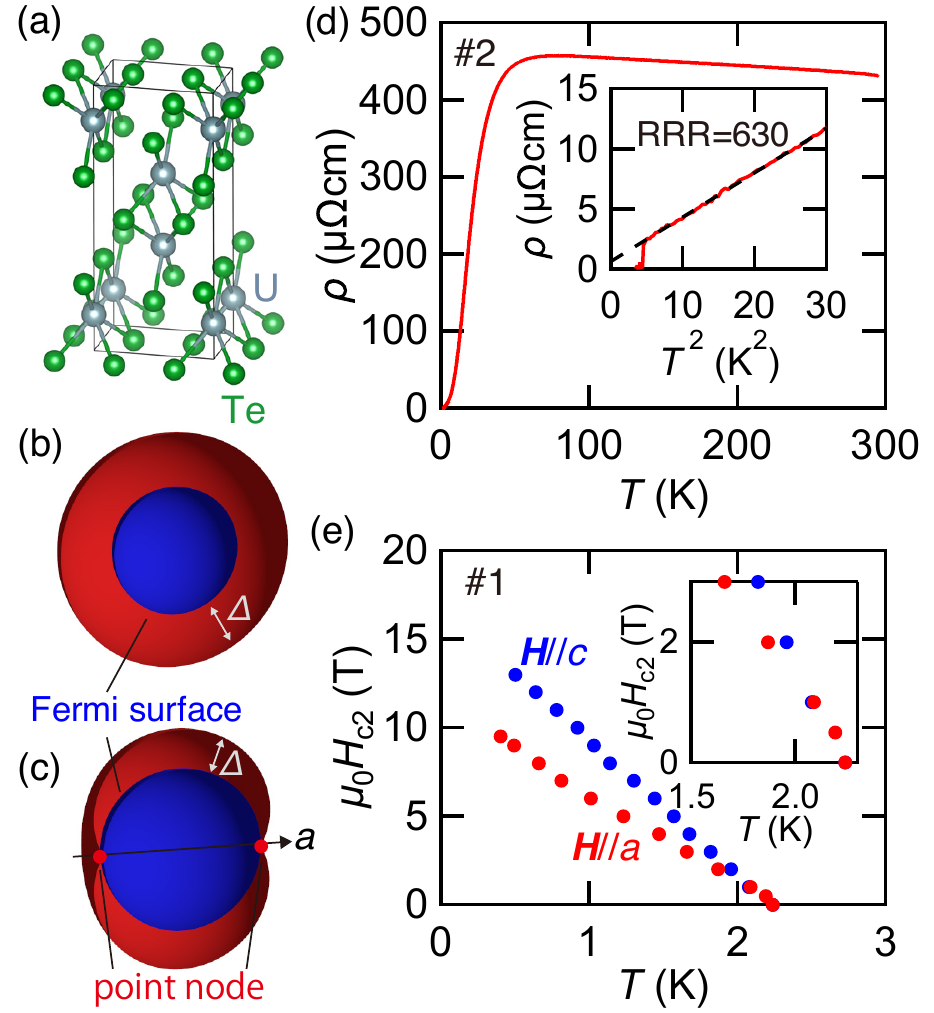}
	\caption{\textbf{Crystal structure, resistivity, and phase diagram of UTe$ _2$.}
		(a) Crystal structure of UTe$ _2$. The gray and green spheres represent U and Te atoms, respectively. (b, c) Structure of the superconducting gap $\Delta$ for $A_u$ (b) and $B_{3u}$ (c) symmetries for spherical Fermi surface (blue sphere). The $B_{3u}$ state has point nodes (red points) along $a$ axis. (d) Temperature dependence of resistivity $\rho$ for sample \#2. The inset shows $\rho$ as a function of $T^2$ at low temperatures. The residual resistivity $\rho_0$ is obtained by a fit to $\rho(T)$ with $\rho_0 + AT^2$ (dashed line). (e) $H$-$T$ phase diagram determined by resistivity measurements of sample \#1. The linear extrapolations to $T=0$ yield the upper critical fields $\mu_0H_{c2}$ of $\sim$ 12\,T for $\bm{H}||a$ and $\sim$ 17\,T for $\bm{H}||c$. The inset shows an enlarged view near $T_c$.
	}
\end{figure}
\begin{figure}
	\includegraphics[clip,width=8cm]{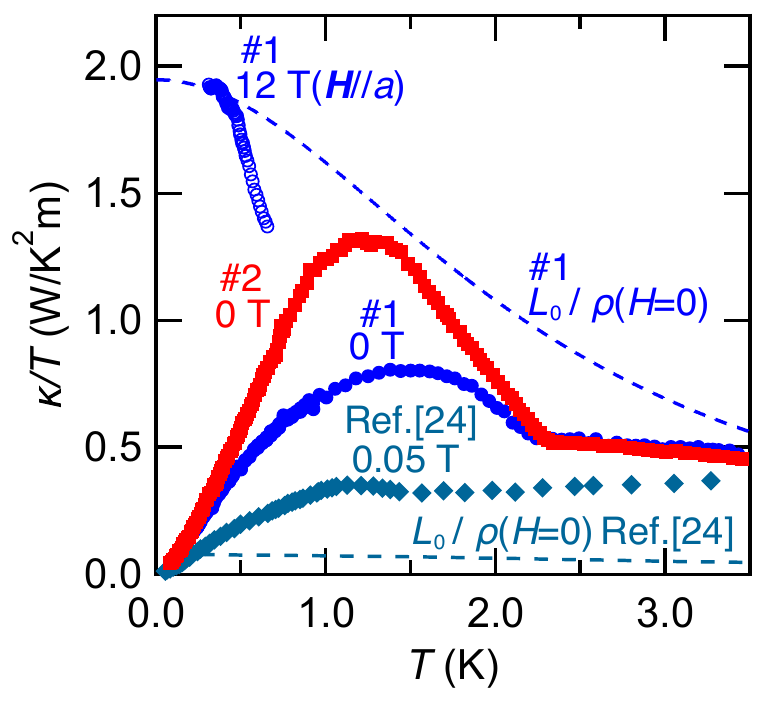}
	\caption{\textbf{Temperature dependence of thermal conductivity of UTe$ _2$.} In zero field, $\kappa/T$ shows a kink at $T_c$, and then increases up to $\sim$ 1.5\,K and $\sim$ 1.2\,K for samples \#1 and \#2, respectively. The blue dashed line represents the electronic contribution estimated by $L_0/\rho(H=0)$ for sample \#1. At low temperatures, $L_0/\rho$ is close to the normal state value of $\kappa/T$ at 12\,T for $\bm{H}||a$ (blue open circles), indicating $\kappa$ is dominated by the electronic contribution. For comparison, we plot $\kappa/T$ at 0.05\,T and $L_0/\rho(H=0)$ for the moderately clean crystal with RRR $= 22$ \cite{PhysRevB.100.220504}.
	}
\end{figure}

\begin{figure*}
	\includegraphics[clip,width=17cm]{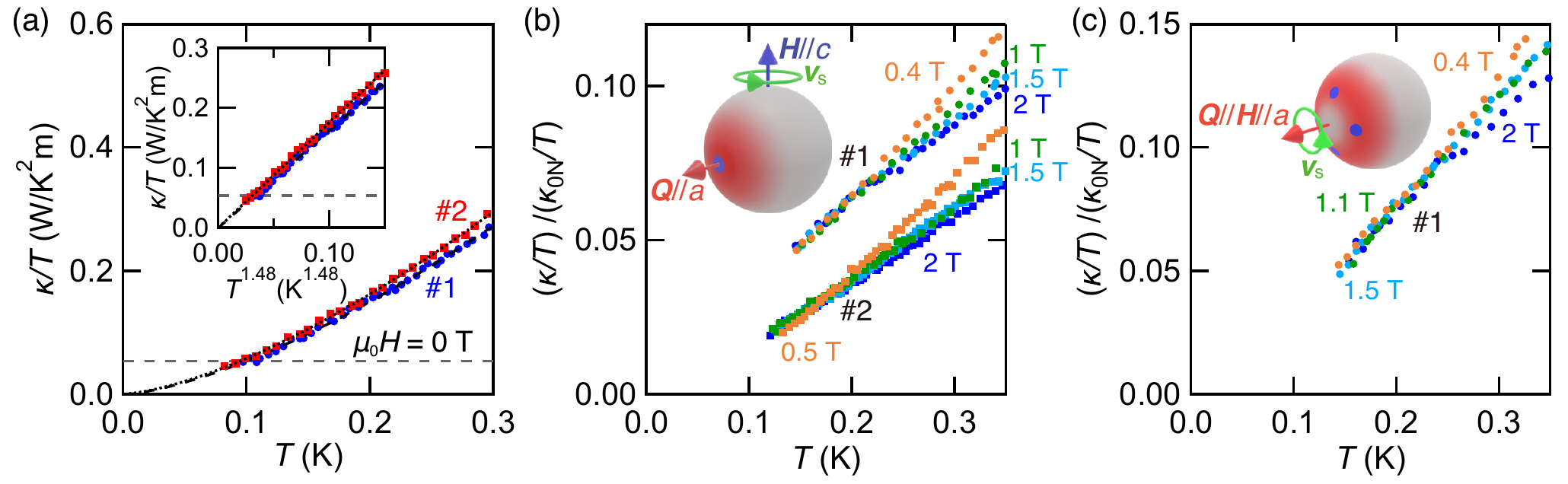}
	\caption{\textbf{Temperature dependence of thermal conductivity of UTe$ _2$ at low temperatures.}
	(a) Thermal conductivity divided by temperature, $\kappa/T$, in zero field for the very-clean (\#1) and ultra-clean (\#2) crystals. The data is extrapolated to $T=0$ by $\kappa/T = \kappa_0/T + AT^{\alpha}$ with $\kappa_0/T \geq 0$ (black dashed and dotted lines), yielding $\kappa_0/T \approx 0$ and $\alpha = 1.48$ for both samples. The inset displays $\kappa/T$ and the fit as a function of $T^{1.48}$. The gray dashed horizontal lines represent the universal constant for line nodes (see text). (b, c) Temperature dependence of $\kappa/T$ normalized by the normal state value in the zero temperature limit $\kappa_\mathrm{N0}/T$ for magnetic fields $\bm{H}$ parallel to the crystal $c$ axis (b) and the $a$ axis (c) with applied thermal current $\bm{Q}||a$. The normal state value $\kappa_\mathrm{0N}/T$ of the very clean crystal for $\bm{H}||a$ is determined by the data at 12\,T (blue open circles in Fig.\,2). $\kappa_\mathrm{0N}/T$ for other configurations is approximated by $L_0 / \rho_0 (H=0)$. As illustrated in the inset, $\kappa/T$ for $\bm{H}||c$ ($\bm{H}||a$) selectively probes the quasiparticles in the red shaded area and is sensitive to the nodes (blue circles) at (around) the $a$ axis. 
	}
\end{figure*}
\begin{figure*}
	\includegraphics[clip,width=17cm]{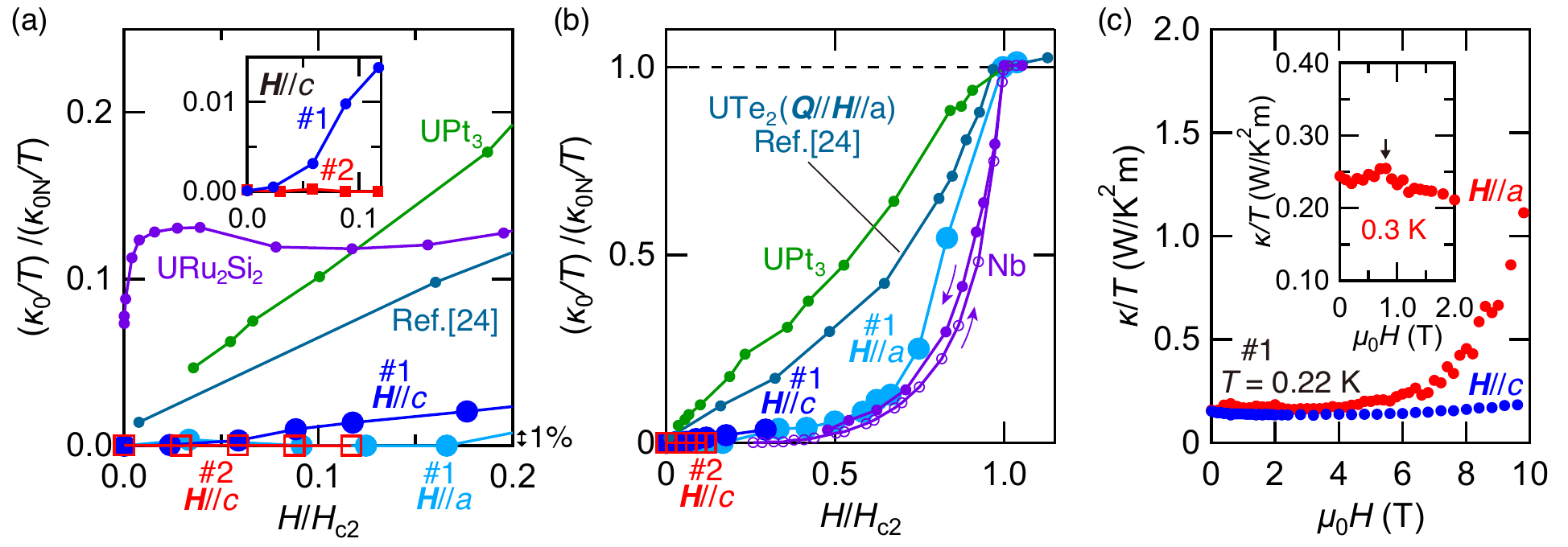}
	\caption{\textbf{Field dependence of thermal conductivity of UTe$ _2$.}
		(a, b) The residual thermal conductivity, $\kappa_0/T$, normalized by the normal state value, $\kappa_\mathrm{0N}/T$, as a function of $H/H_\mathrm{c2}$. For both $\bm{H}||c$ and $\bm{H}||a$, $\kappa_0/T$ is less than 1\% of $\kappa_\mathrm{0N}/T$ up to $H/H_\mathrm{c2}=0.09$, which provides evidence for the absence of any types of nodes at or around the $a$ axis. The inset shows an enlarged view in the low field region for $\bm{H}||c$. For comparison, we plot data for other uranium superconductors UPt$ _3$ \cite{suderow1997thermal} and URu$ _2$Si$ _2$ \cite{PhysRevLett.99.116402} (a) and typical $s$-wave superconductor Nb \cite{lowell1970mixed} and nodal superconductor UPt$ _3$ \cite{suderow1997thermal} (b). We also plot the data for the moderately clean crystal with RRR $= 22$ for $\bm{Q}||\bm{H}||a$ \cite{PhysRevB.100.220504}. (c) Field dependence of $\kappa/T$ at 0.22\,K for $\bm{H}||c$ and $\bm{H}||a$ for sample \#1. The inset displays the field dependence of $\kappa/T$ at 0.3 K for $\bm{H}||a$. The peak anomaly is found at $\sim$ 0.8\,T (black arrow).
	}
\end{figure*}

%
\end{document}